\documentclass[12pt,preprint]{aastex}

\usepackage[caption=false]{subfig}

\slugcomment{}

\shorttitle{Secondary eclipse detection}
\shortauthors{Burton et al.}

\begin{document}

\title{$z$'-band ground-based detection \\
of the secondary eclipse of WASP-19b}

\author{J. R. Burton\altaffilmark{1}, C. A. Watson\altaffilmark{1}, S. P. Littlefair\altaffilmark{2},
V. S. Dhillon\altaffilmark{2}, \\N. P. Gibson\altaffilmark{3}, T. R. Marsh\altaffilmark{4}, 
D. Pollacco\altaffilmark{1}}

\email{jburton04@qub.ac.uk}

\altaffiltext{1}{Astrophysics Research Centre, Queen's University Belfast, Belfast, BT7 1NN, UK}
\altaffiltext{2}{Department of Physics and Astronomy, University of Sheffield, Sheffield, S3 7RH, UK}
\altaffiltext{3}{Department of Physics, University of Oxford, Oxford, OX1 3RH, UK}
\altaffiltext{4}{Department of Physics and Astronomy, University of Warwick, Coventry, CV4 7AL, UK}

\begin{abstract}
We present the ground-based detection of the secondary eclipse 
of the transiting exoplanet WASP-19b. The observations were made in the 
Sloan $z$'-band using the ULTRACAM triple-beam CCD camera mounted 
on the NTT. The measurement shows a 0.088$\pm$0.019\% eclipse depth, 
matching previous predictions based on $H$- and $K$-band measurements.
We discuss in detail our approach to the removal of errors arising due to systematics
in the data set, in addition to fitting a model transit to our data. This fit returns
an eclipse centre, $T_0$, of 2455578.7676 HJD, consistent with a circular orbit.
Our measurement of the secondary eclipse depth is also compared to model
atmospheres of WASP-19b, and is found to be consistent with previous measurements
at longer wavelengths for the model atmospheres we investigated.
\end{abstract}


\section{Introduction}

Transiting exoplanetary systems provide an excellent opportunity to measure 
the physical properties of exoplanets, and can allow for the atmospheric composition and 
structure to be investigated. The primary transit (where the planet occults the star), 
combined with radial velocity measurements mean key planetary parameters, 
such as the radius and mass, can be inferred for the exoplanetary system. 
The secondary eclipse, where the planet passes behind the sky-projected 
disc of the star, allows for the direct detection of flux from the exoplanet, 
meaning properties of the planet can be directly measured, instead of 
derived from the transit light curve or the radial velocity. For example, 
observations of the secondary eclipse provides information on the temperature 
(\citealt{knutson07}) and atmospheric constituents (e.g. \citealt{burrows05}, 
\citealt{burrows06}) of the exoplanet, and has been a powerful tool into the study 
of hot-Jupiters and their atmospheres. Previous work on secondary eclipses has 
mostly been carried out from space-based platforms (notably the Spitzer space 
telescope, e.g. \citealt{laughlin09}, \citealt{deming10}), but recent work has been 
focused on obtaining secondary eclipse detections from the ground (e.g. 
\citealt{demooj09}, \citealt{alonso10}, \citealt{zhao11}). 

The transiting exoplanet WASP-19b (\citealp{hebb10}) has one of the shortest-known 
orbital periods of 0.79 days. The level of irradiation incident on the planetary 
surface, coupled with poor heat redistribution (\citealt{fortney08}) makes WASP-19b 
one of the hottest known transiting exoplanets. In addition to this, systems with 
short periods such as these are subject to intense tidal forces (e.g. \citealt{leconte11}), 
meaning WASP-19b is an extremely interesting case for both atmospheric 
composition and structure models. The secondary transit of WASP-19b has 
previously been observed from the ground in the $H$- and $K$-bands using the 
HAWK-I (High Acuity Wide-field K-band Imager) instrument mounted on the VLT 
(\citealt{anderson10}, \citealt{gibson10}). The secondary eclipse depth in these 
studies was found to be 0.259$\pm$0.045\% and 0.366$\pm$0.072\% respectively, 
corresponding to a dayside brightness temperature of $\sim$2500K in both cases. 
From this, the authors concluded that there was poor heat redistribution from dayside 
to nightside, and demonstrated from the phase offset of the eclipse centre that the orbit 
of WASP-19b is consistent with a circular orbit. 

In this paper, we present a ground-based secondary eclipse observation of 
WASP-19b in the Sloan $z$'-band (centred on 909.7nm). We discuss the data acquisition 
and reduction, along with the limitations associated with our ground-based observations, 
and potential follow-up work. This is only the 3rd $z$'-band detection of an 
exoplanet secondary eclipse from the ground, the first being  OGLE-TR-56b 
(\citealt{sing09a}), and the second being WASP-12b (\citealt{lopez10}).

\section{Observations}

On the 17th January 2011, the $m_v$=12.3 star WASP-19 was observed 
from UT 01:56-09:02. It was observed simultaneously in the Sloan-$u$', $g$' and $z$' bands 
using the high speed CCD camera ULTRACAM (see \citealt{dhillon07} for a description) on 
the European Southern Observatory (ESO) 3.5m New Technology Telescope (NTT) based 
at La Silla, Chile. The frame-transfer capability of ULTRACAM means that there is negligible
deadtime ($\sim$25ms) between exposures, maximising the efficiency of 
observations. This high-efficiency mode allows for the chip to be read out while the 
next data frame is being exposed, meaning the CCD can obtain thousands 
of images per night, in addition to exposing in three bands at once.

In order to maximise the precision during the observations, we opted to defocus the 
telescope
(see e.g. \citealt{southworth09} for an in-depth description of the rationale and techniques). 
This 
was further assisted thanks to ULTRACAM's design allowing for approximately 600 twilight 
flat-field-frames to be taken per night, giving an excellent characterisation of the 
pixel-to-pixel variations of the CCD. A number of steps were also taken to ensure the 
observations were as free from systematics as possible whilst at the telescope. A 
previous observing run using ULTRACAM on the NTT had discovered variable 
vignetting over the edge of the chip as a result of the positioning of the guide probe. 
In order to ensure this did not happen during our exposures, the guide probe position 
was carefully selected such that vignetting was avoided throughout the observations, 
and positioned out of the beam when obtaining flat fields.

While acquiring flat-field exposures, the telescope was spiralled so that any 
stars present did not remain on the same pixel over consecutive images. Sky flats were also 
taken near the zenith where the sky brightness as a function of altitude varies the least,
minimising gradients across the chip. Over the course of the data collection, utmost care 
was taken for the target to remain on the same position on the chip. We carefully picked 
the guide star to avoid vignetting by the guide probe, continuously monitored both the x- 
and y-position of the target on the ULTRACAM CCD in real time, and manually corrected 
any drift when necessary. Doing this allowed for the drift of the target to be less than one 
pixel throughout the majority of the night (the mean motion of the target over all the frames 
was calculated to be 0.47 pixels or 0.16"). The telescope was defocussed so 
that the objects' point spread function (PSF) had a full-width at half-maximum (FWHM) of 4-5". 
Since the WASP-19 field has a number of objects at a relatively small separation from the 
target, 4-5" defocus was found to be the maximum possible before blending with background 
objects might have become an issue. Observation conditions over the night were very good, 
with seeing remaining at $\sim$1" over the course of the exposures. In addition to this, we 
also used the simultaneously-recorded $g$'-band to monitor the surface activity of the star, 
again, in real time. Doing this also allowed us to analyse the $z$'-band light curve for systematics 
(since the secondary eclipse signal of the planet is too faint to see in the $g$'-band), and since 
the spot/photosphere contrast is higher in the $g$'-band compared to the $z$'-band, any features 
on timescales longer than the secondary eclipse could also be monitored with a high degree of precision.

\section{Calibration}

4524 data frames were obtained per filter over the course of the night, each with an
exposure time of $\sim$5.7s. We obtained approximately 120 minutes of pre-ingress 
data, and 90 minutes of post-egress data, meaning the out-of-eclipse portion 
of the lightcurve would be well-characterised (the predicted secondary eclipse duration 
for WASP-19b is 92.4 minutes as a comparison). Once the data had been obtained, 
photometry was performed first by de-biasing and applying flat-fields in the usual way. 
The ULTRACAM pipeline\footnote[1]{http://deneb.astro.warwick.ac.uk/phsaap/ultracam} 
was used to create apertures around the target and six bright comparison objects in the 
field of view (see figure 1). The apertures were set to a fixed radius, and the source position 
was tracked using a moffat fit on each frame. We also ran the entire data extraction using a 
Gaussian fit in order to ensure that different tracking methods did not affect how the sources 
were tracked over the course of observations, and there was no difference in the resulting lightcurves. 
The object aperture was set to a radius of 18 pixels. The inner sky aperture, used for sampling the 
background level, was set to a radius of 26 pixels and the outer sky aperture was set to a radius of 
100 (we note here that a number of different aperture sizes were trialled, and as long as a 
reasonable sky background was sampled, this did not affect the target or comparison lightcurves). 
Since a number of background objects were in this aperture, these were masked out in order 
to remove them from the sky background estimation. Differential photometry was then performed on 
the target by dividing by the average flux of the six reference stars in the field. For the $g$'-
band, only 5 comparisons were used since one of the reference stars became too heavily 
blended with 2 background objects. The $u$'-band had only 2 visible comparison objects, 
as the majority of the stars were too faint to be detected at this shorter wavelength. We do 
not present any of the $u$'-band data due to the lack of suitable comparison objects and 
hence, the noise present in the lightcurve.

\begin{figure}[!h]
\epsscale{0.8}
\plotone{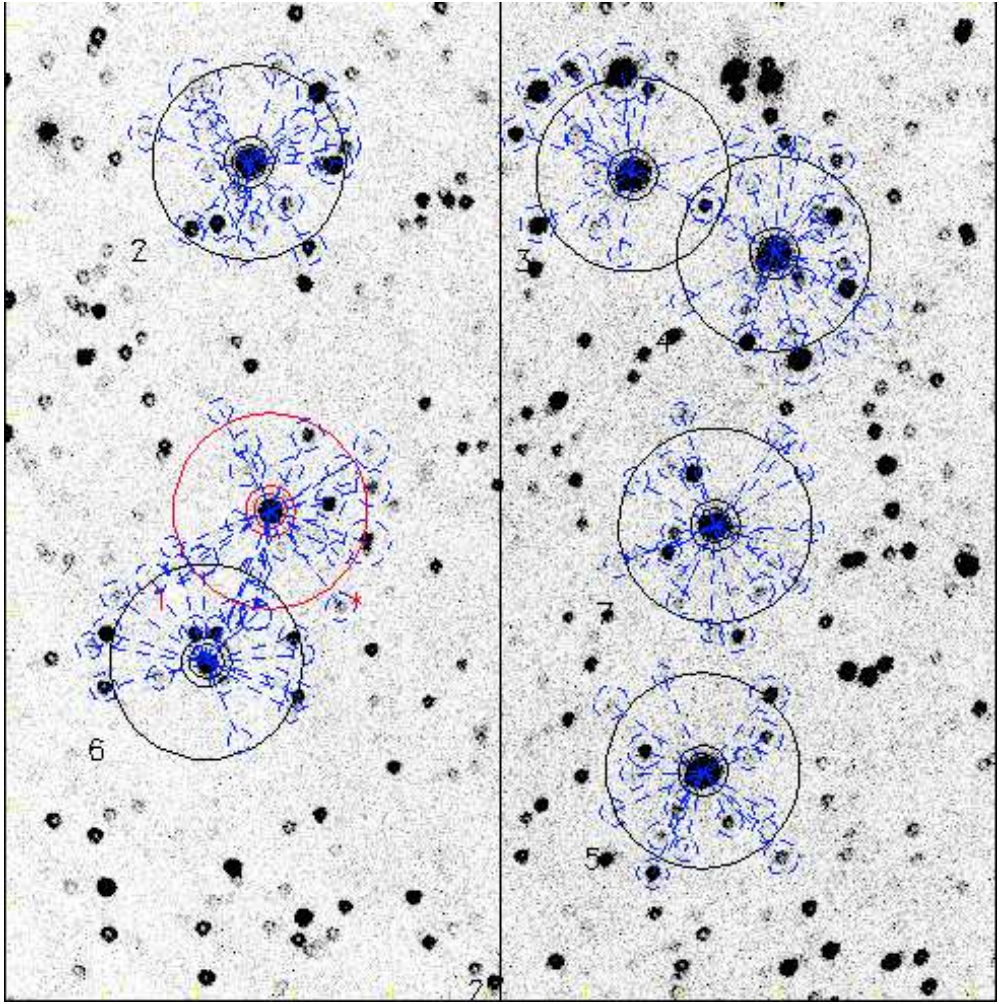}
\caption{$z$'-band data frame showing the WASP-19 field, after the colours have been inverted 
for the sake of clarity. North is directed upwards and east is to the left in this image. The target is 
located in the red (light grey) aperture (left-hand panel, middle object), and the six comparisons 
used are shown in the remaining apertures. These indicate the sky and target apertures used for 
the $z$'-band data reduction. The smaller dashed apertures (blue/dark grey) indicate objects 
which have been masked out from the sky background. The field-of view (FOV) in our case measures 6 arcminutes.\label{fig1}}
\end{figure}

\section{Data Analysis}

Given the challenges of obtaining a $\sim$1 mmag secondary eclipse detection, 
the majority of our data analysis has been concentrated on characterising and 
removing systematics that arise as a result of our ground-based observations 
as well as identifying any correlations which appear to be present. This section 
focuses on our method of identifying and removing systematic errors using a 
`weighting' method in the comparison lightcurve, along with the process of 
decorrelation to ensure the target lightcurve is as free from systematics as 
possible. We also describe the fitting of a model transit to the data to infer the 
secondary eclipse depth and error.

\subsection{Systematics} \label{bozomath}

\begin{figure}[!h]
\epsscale{1.0}
\plotone{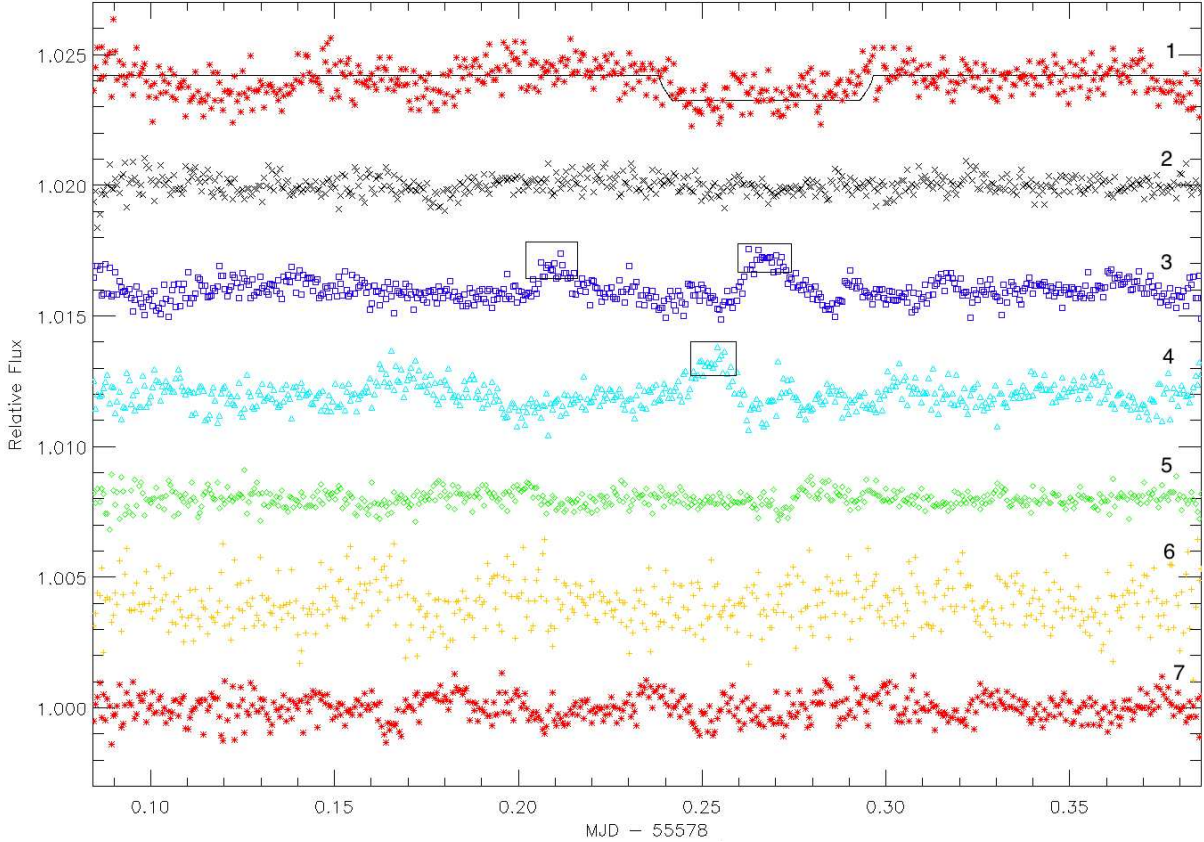}
\caption{Normalised lightcurves (using a 2nd-order polynomial fit) of the target (1) and the six 
comparisons (2-7) corresponding to the objects shown in figure 1 after dividing through by the 
total comparison lightcurve. The model transit (black line) is our initial Mandel-Agol fit (see section 4.3). 
The 3 boxed sections in comparisons 3 and 4 indicates regions where systematic `spikes' flagged 
by our analysis (see text) were high. We note here that these data points have not been rejected 
outright, but are merely indications of regions in which our systematic analysis flagged a number 
of systematic errors. At these times, these comparisons are rejected and the remaining weighting factors 
are adjusted to remove this effect from the total lightcurve.\label{fig2}}
\end{figure}

Since our observations were made from
the ground, systematics may be appreciable compared to the signal from any 
potential secondary eclipse detections. As is normal when carrying out differential 
photometry, we obtained the final WASP-19 lightcurve, $L(t)$, at time $t$ using the 
following formula:

\begin{equation}
L(t) = \frac{T(t)}{a_1(t)C_1(t)+a_2(t)C_2(t)+...+a_n(t)C_n(t)} ,
\end{equation}

\noindent where $T(t)$ is
the raw flux from the target, $C_n(t)$ is the flux from comparison star `n', and $a_n(t)$ 
is the weighting factor of comparison `n'. When carrying out differential photometry, 
the weighting factor is normally set to 1. In our analysis, however, we allow for the 
weighting factor to be switched to 0 depending on whether we identify a systematic 
in the raw lightcurve of the comparison object. The other weighting factors for the 
remaining comparisons are then adjusted together to take into account the missing 
flux due to the rejected comparison contribution. In our analysis, if the contribution 
from the comparison showed a sustained run of points which lay more than 1$\sigma$ 
from a local mean, we rejected these points as due to a systematic -- essentially if the 
$a_n(t)$ value in equation (1) was abnormally high or low for a number of consecutive 
data points. We defined a systematic region as any run of 6 or more consecutive points, 
all of which lay $>$1$\sigma$ above or below the mean contribution for that comparison. 
Statistically, the chances of this happening are $<$0.001\%. Once this analysis had been 
carried out for all comparisons, the points which had been rejected due to systematics were 
flagged in the total comparison lightcurve, and the weighting factors of the remaining comparisons 
adjusted such that this systematic was removed. Extreme care was taken to only adjust for points which 
showed systematic errors so as not to falsely idealise the comparison lightcurve. 
Figure 2 shows the lightcurves of WASP-19 (1), along with each of the comparisons 
(2-7), corresponding to the numbered annuli in figure 1. The boxed sections indicate 
regions which contained a high concentration of points which were flagged as systematics in our 
analysis. The majority of points rejected appear to be in comparison objects 3 and 4. As can be 
seen from figure 1, these objects have a large number of background objects, some of which are 
of appreciable brightness, and appear near the edge of the right-hand side of the CCD. This may 
go some way as to explaining why these objects show an increase in systematics at these times, 
as some background stars may have leaked into the sky annulus over these observations. In 
addition to this, some level of vignetting may remain, these two objects being the most susceptible 
to being affected due to their close proximity to the upper edge of the CCD and their close proximity 
to each other. The rejected points can then be removed, the contributions of the remaining comparisons 
increased, and the lightcurve of the target improved. Upon comparison of the pre- and post-correction lightcurves of WASP-19, 
little difference is noticeable in a significant number of data points once we remove any points flagged as a systematic. Even when reducing the
number of consecutive 1-$\sigma$ outliers to 5 or even 4, the difference is minimal. This indicates
that the WASP-19 lightcurve is robust against the systematics that we have identified arising in 
the comparison lightcurves. This is probably due to the fact that we have enough good comparison 
objects in the field to remove any significant systematic errors before our analysis.

\subsection{Correlations} \label{bozomath}

The presence of a correlation or anti-correlation between the flux from the target 
and another parameter (such as position of the star on the chip, airmass, sky 
background etc.) may indicate that any features present in the lightcurve 
(e.g. a systematic mimicking a secondary eclipse) could be related to this correlation. 
During our data analysis, all parameters which we were able to quantify were
thoroughly tested for the presence of correlations, and if there was evidence 
of this, a decorrelation could be applied to check if the secondary eclipse feature 
remained. We thoroughly tested the following for correlations against the flux 
from the target; x-position, y-position, mean position (i.e. simultaneous x- and y-position on the CCD), 
sky background, airmass (altitude), azimuth and seeing (the FWHM of the target's PSF), 
in addition to investigating the relationship between the wavelength bands themselves 
(i.e. a direct comparison between the $z$'-band and $g$'-band). The resulting plots 
(see appendix) indicate that the strongest correlation was between the $z$'-band and 
sky background, where there appears to be almost a 2-component trend, where the lower 
sky counts appear to correlate with higher $z$'-band fluxes. This is shown in figure 3, 
in which the points where the planet is predicted to be in secondary eclipse are highlighted. 
All of the other parameters which we tested for showed either no significant correlations or 
showed low-level correlation which disappeared after the sky correction. This analysis was 
also applied to each of the comparison objects in turn, in order to check for the same correlations.

\begin{figure}[!h]
\epsscale{1.0}
\plotone{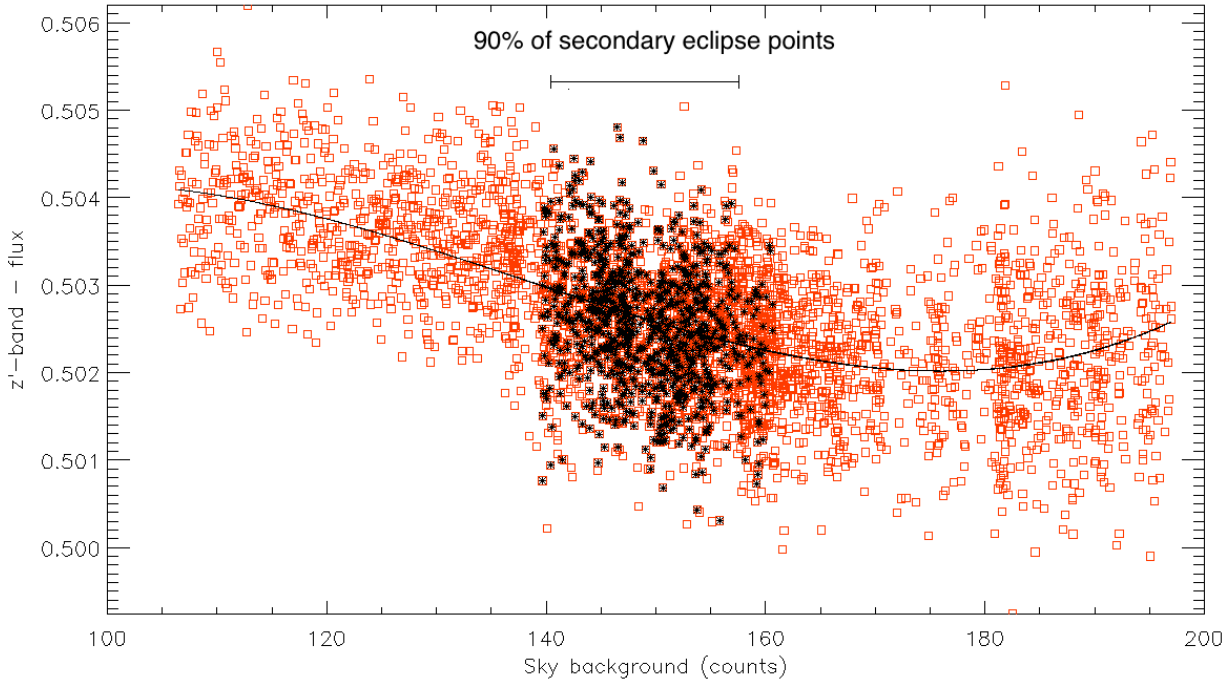}
\caption{Sky vs. target $z$'-band, along with the 3rd-order polynomial we used to fit the trend. Points where the planet is predicted to be in secondary 
eclipse have been highlighted in black (asterisks). The horizontal bar indicates the 
range in sky background counts over which 90\% of the secondary eclipse points 
lie.\label{fig3}}
\end{figure}

Plotting the lightcurve of the sky as a function of time reveals a number of features 
present in the sky background over the course of observations. Figure 4 shows a 
number of features pre-ingress (i.e. before 0.21MJD) and post egress - when we 
entered dawn twilight. It can also be seen that there are no major features present 
in the sky background during the predicted secondary eclipse ($\sim$0.23-0.3MJD), 
other than a general slope. This indicates that the sky background was well-behaved 
and free of any major features which may impact the target lightcurve 
during secondary eclipse. We note here that all correlations and tests for 
systematics have been cut off before 0.35MJD due to the sharp increase in sky 
background (due to the sunrise) after this time. It is also important to state that 
the object transited across the meridian just after egress (0.3MJD).

\begin{figure}[!h]
\epsscale{0.9}
\plotone{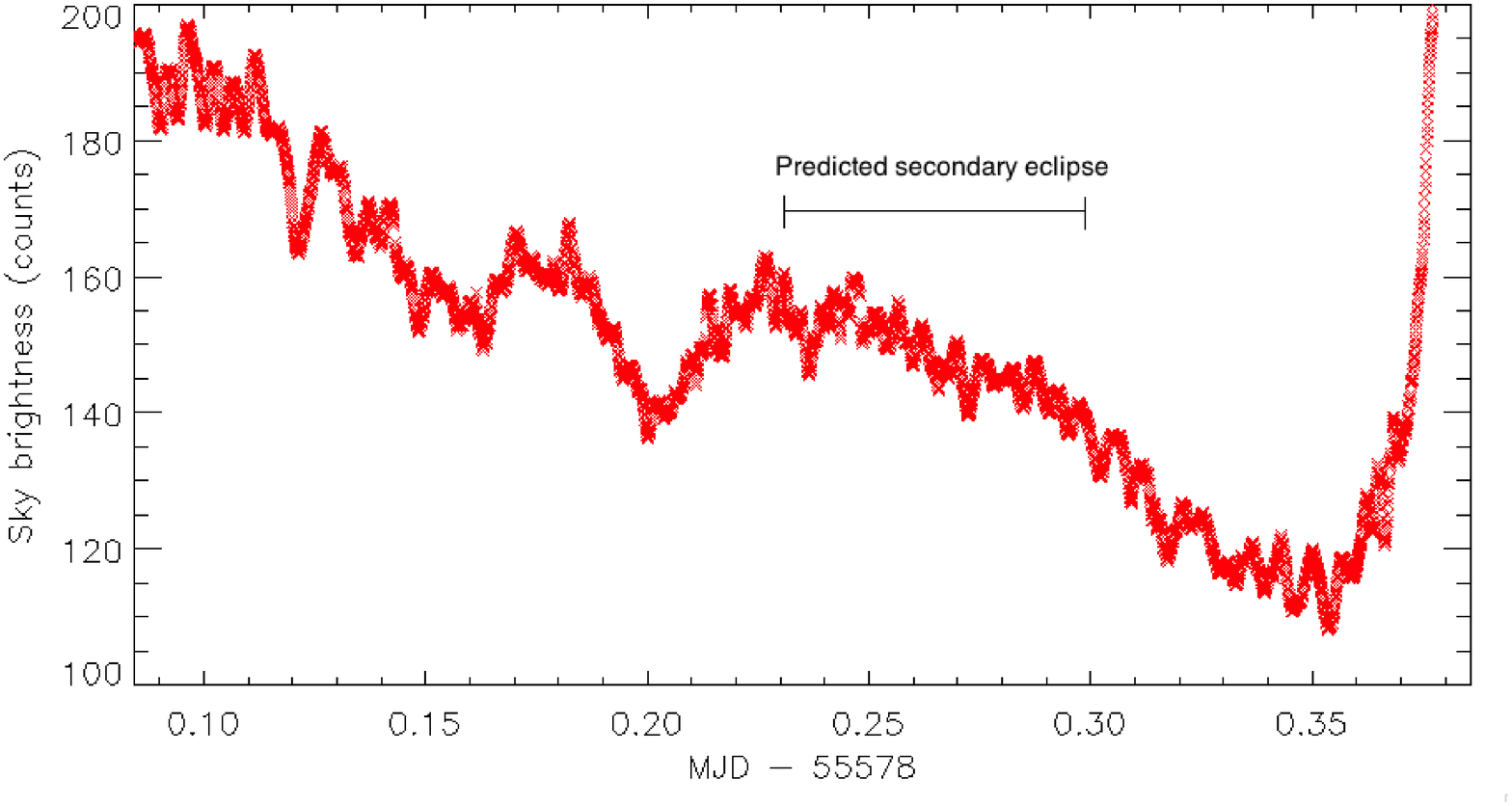}
\caption{$z$'-band sky background over the course of observing. Note the numerous features
in the lightcurve, especially the points between 0.15-0.22 MJD (i.e. just before the planet enters
secondary eclipse). Altering these points had a direct effect on the target lightcurve, indicating
the care which must be taken when dealing with the sky background. \label{fig4}}
\end{figure}

In order to adjust for the sky background, we first modelled the correlation using
a polynomial fit. We then found the average value of the $z$'-band flux, 
and used this in our polynomial formula to give an adjustment factor for each
data point. We then simply divided our target lightcurve by this adjustment factor
in order to remove the correlation with sky. In addition to the polynomial fit, a number
of other methods were also used to model the correlation (e.g. spline fit, linear fit). 
However, an important factor was to be able to smoothly vary from one
component into the other, something which was fairly difficult to satisfactorily achieve 
using a spline fit. We opted to use a 3rd-order polynomial fit based on this, combined with the 
fact that the polynomial did remove the effects of sky brightness to an excellent degree.
We tested higher-order polynomials (4th and 5th term), but the fit provided by these
resulted in an over-correction, due to the spread of the points in figure 4. With higher order fits,
we also found that the points in the lightcurve where the planet is in egress were affected,
as these points lay where the sky brightness correlation changed rapidly. This meant we
were extremely cautious both when applying our correlation correction to the target lightcurve as
well as fitting the model to the secondary eclipse.
We also tested how changing the polynomial at the lower sky brightness affected the
target lightcurve. We discovered that altering the polynomial below 140 counts 
altered the post-egress data points, noting that the secondary eclipse feature was
not affected by altering the order of the polynomial. This can be seen in figure 3,
where the points which correspond to the secondary eclipse (black points) are all
located in the main concentration of data points (i.e. 140-170 counts sky background).
The result of this correction is shown in figure 5, below.

\subsection{Transit fitting} \label{bozomath}

In order to fit the lightcurve, a simple model transit was generated using the technique outlined by 
 \citet{mandel02} of an opaque disc occulting a second disc. For this, we assumed zero limb 
darkening for the exoplanet. Once the systematic and decorrelation analysis had been 
carried out, the best fit to this secondary eclipse using the Markov-Chain-Monte-Carlo 
(MCMC) technique was found, the approach of which is described by \citet{cameron07}. 
Initially, we made the assumption that the planet is in a circular orbit around the star, 
and in addition the secondary eclipse times for the night we observed were fixed and accurate 
(in the second step of our fitting process, we allowed the transit duration, and the eclipse centre
time -- $T_0$ -- to vary).
In order to obtain the correct depth and associated error, we used the Transit Analysis
Package (TAP)\footnote[2]{http://ifa.hawaii.edu/users/zgazak/IfA/TAP.html} to model our single 
transit. This is a GUI written for IDL which allows for MCMC analysis to be performed with a 
number of user-defined system parameters (mass ratio, period, eclipse centre, eccentricity and inclination of the system). 
Since we were assuming the feature present in the lightcurve was the actual secondary 
eclipse, the system parameters were fixed based on the values given by 
\citet{hellier11}. Our priority in this analysis was solely to obtain the eclipse depth and errors 
(termed `white noise' in the TAP interface), which meant fixing the system parameters was 
justified (in the same manner as \citealt{gibson10}). 
The TAP returned an eclipse depth and associated error of 0.1$\pm$0.02\% (1.0$\pm$0.2 mmag).
We then double-checked the results from the TAP using the secondary eclipse fitting routines 
of \citet{gibson10}. Since these have had a legacy of successful secondary eclipse fitting 
(for the same planetary system), these routines were the preferred choice to fit 
the eclipse depth, transit times and error. Upon employing the Gibson routines, the system parameters used 
were, again, from \citet{hellier11}, and fixed in the MCMC run. 
After running the routines of \citet{gibson10} several times whilst fixing the transit duration, the eclipse 
depth was found to be 0.081$\pm$0.018\% (0.81$\pm$0.18 mmag), in close agreement with the result 
from the TAP. In the case where we allowed the transit duration to vary, however, we returned an eclipse 
depth of 0.088$\pm$0.019\% (0.88$\pm$0.19 mmag). This was due to a shorter transit time allowing for fewer 
points to lie above the model, resulting in a better fit of the model to the data. When we isolate the secondary eclipse
by opting to fit the transit between 0.22 and 0.32MJD from the same initial model, we return the same eclipse depth
with a slightly reduced error - 0.088$\pm$0.015\%. This is due to the fact that we are analysing fewer points which
lie predominantly above the normalisation level just before and after the secondary eclipse feature.
Out of the two eclipse depths 
we return, we have opted to draw our conclusions from the analysis where we allowed the transit duration 
to vary, since our single night's data cannot put constraints on the transit duration and $T_0$ value with too high a level of 
confidence. We have also opted to draw our conclusions
from the total lightcurve duration fit, as this gives a better indication of the errors present in the out-of-eclipse baseline.
Table 1 shows the parameters which resulted from the MCMC analysis, and the final secondary 
eclipse is shown in figure 5.

\begin{center}
Table 1: MCMC parameters
\end{center}
\vspace{-20pt}

\begin{table}[!h]
\centering
\renewcommand{\tabcolsep}{1.cm}
\renewcommand{\arraystretch}{1}
\begin{tabular}{l|ccr}
Parameter & Symbol & MCMC values & Unit \\ 
\hline
Eclipse depth & D & 0.088$\pm$0.019 & \% \\
Transit centre & $T_0$ & 2455578.7676$\pm$0.0039 & HJD \\
\end{tabular}
\end{table}

\vspace{-15pt}
\begin{center}
\footnotesize{Parameters from our final MCMC fitting routine.}
\end{center}

\section{Results}

\begin{figure}[!h]
\epsscale{1.0}
\plotone{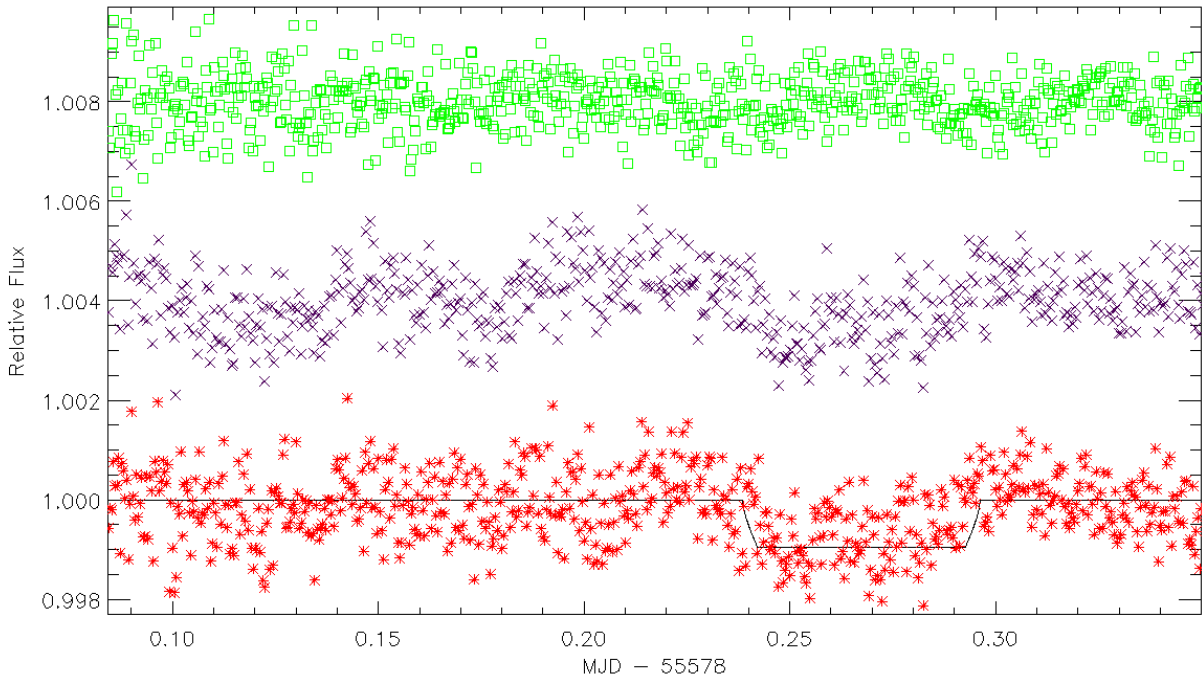}
\caption{Target lightcurve pre- and post-decorrelation, normalised using a second order polynomial. 
The uncorrected $z$'-band is shown in black (central), the $z$'-band data after decorrelation and systematic analysis is shown in red (lower), and the green (top) points show the $g$'-band data with 
no feature at the time of secondary eclipse (as expected). The solid black line is our best-fit model 
of the secondary eclipse to the data.\label{fig5}}
\end{figure}

The target lightcurve is shown in figure 5. As can be seen from the uncorrected lightcurve 
(central data points), at time 0.1 to 0.14MJD, there appears to be a feature of a similar depth and 
duration as the secondary eclipse. However, once we apply our decorrelation to this, 
this feature disappears, and more importantly, the feature at 0.23 to 0.3MJD (the 
approximate predicted ingress and egress times for the secondary eclipse) does not change. In 
addition to this, our analysis of the systematics on each of the comparison objects 
shows little variation once we apply our technique of `weighted comparison contribution',
indicating that whatever systematics remain are a feature of the target. During our decorrelation, 
the majority of features in the out-of-eclipse portion of the lightcurve were 
removed. After detrending, the secondary eclipse still remains at the same 
epoch, and appears to remain at a constant depth throughout the decorrelation and 
systematic removal process. As can be seen from figure 5, a number of the features present pre-ingress appear to 
be reduced once our sky correlation correction is applied. However, the secondary 
eclipse feature from $\sim$0.23-0.30MJD remains and is robust against both the systematic 
analysis and sky decorrelation; an indication that the feature which occurs at the predicted 
secondary eclipse ingress and egress times is a feature inherent to the target for this 
data, and not due to systematics of the comparison objects or correlations with the sky 
background. 
The systematic analysis and decorrelation method was also 
applied to the $g$'-band, and the resulting lightcurve is also shown in figure 5 (top data points). As 
with the $z$'-band, the strongest correlation for the $g$'-band was with sky brightness 
(see appendix). We used the same method to remove this trend for both bands. The sky 
background correlation was much more apparent in the $g$'-band, and once this was 
removed, the remaining correlations with position, altitude etc. were significantly reduced, 
indicating the amount of influence the sky background has on the data set. As can be seen 
from figure 5, the feature which appears at the correct ingress time for the secondary eclipse 
in the $z$'-band does not appear in the $g$'-band, indicating the feature is unique to the $z$'-band. 
This is further evidence of the detection of the secondary eclipse, as the eclipse should be too faint to 
detect in the $g$'-band, consistent with our results. When we apply a fit to the $g$'-band data using 
the same system parameters as with the $z$'-band lightcurve, the routines return an eclipse depth of 
0.001$\pm$0.028\%, negligible in comparison to the $z$'-band.
From table 1, the $T_0$ parameter is returned to be 2455578.7676 HJD from our analysis when we allow the 
transit duration to vary. In comparison to the transit ephemerides provided by \citet{hebb10}, the predicted time 
of mid-eclipse, assuming a circular orbit is given to be 255578.76962 HJD. The shorter transit duration we obtain
could be explained by the decrease in flux post-egress, possibly due to the observations transiting the meridian just after
egress. This $T_0$ value corresponds to a phase of 0.490$\pm$0.0047, consistent with the results of \citet{gibson10} 
indicating that WASP-19b is in a circular orbit. We note here that while our formal statistical error
from our MCMC fit is 0.0047, due to the noise present in the lightcurve as well as any systematics
which remain unaccounted for, this error is likely to be higher.
Since this is based on a single night's data, follow-up observations will undoubtedly further constrain 
the time of mid-eclipse, even observations made post-ingress will be helpful in constraining this parameter 
in addition to the eclipse times. When we take into account the error on the $T_0$ value, our transit duration 
agrees to the predicted value to $\sim$1.5$\sigma$, a result which again can be improved upon with 
additional observations. 

\begin{figure}[!h]
\epsscale{0.6}
\plotone{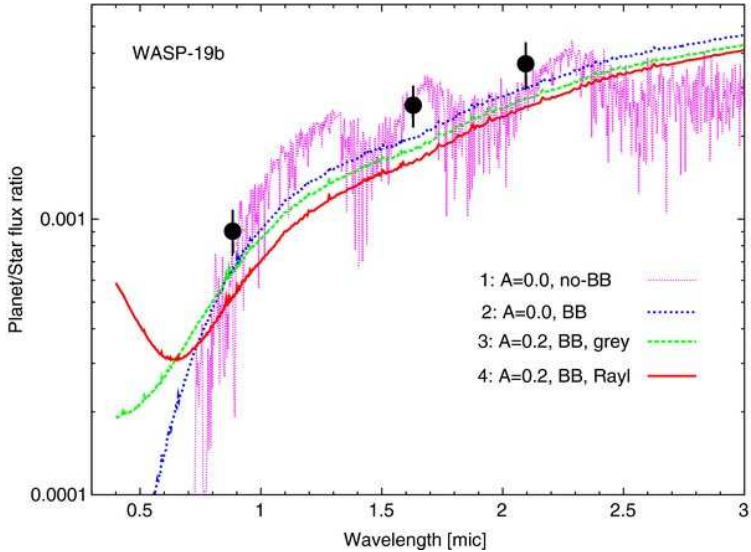}
\caption{Various atmospheric models of the planet WASP-19b (from Budaj, 2011). Model 1 
(dashed purple line) is the non-blackbody model based on non-irradiated atmospheres. 
Model 2 (dashed blue line) is a blackbody model with zero albedo. Model 3 (dashed green line) 
is a grey albedo model with A$_B$ = 0.2. Model 4 (solid red line) is a non-grey albedo model
with Rayleigh scattering and the same Bond albedo (A$_B$ = 0.2). The right-hand 
and central black points are the $K$- and $H$-band measurements from \protect\citet{gibson10} 
and \protect\citet{anderson10} respectively. The left-hand point indicates the depth of the secondary 
eclipse signature from our ULTRACAM + NTT $z$' -band observations. \label{fig6}}
\end{figure}

The near-infrared provides an important window for constraining the atmospheric 
pressure-temperature profile at depth. Atmospheric models generally indicate that 
the main opacity source is due to water vapour which is particularly prominent in 
the mid-infrared. However, in the near-infrared it has been proposed (e.g. 
\citealt{fortney08}, \citealt{burrows08}) that an opacity window appears which 
allows the atmosphere to be probed more deeply to gas lying at higher pressure. 
For exoplanets with temperature inversions in their atmospheres, the near-infrared 
emission should appear weak compared to redder observations since such planets 
will feature a relatively cooler atmosphere at depth compared to the hotter upper 
atmosphere. For planets with no temperature inversion, the gas at depth is also hot, 
resulting in a much reduced contrast in emission properties as one scans across 
multiple wavelengths. Thus, $z$'-band observations may provide constraints on 
the pressure-temperature gradient in the atmosphere, as well as an insight into the 
opacity mechanisms in operation (e.g. \citealt{croll08}). An eclipse depth of 
0.088$\pm$0.018\% allows us to begin to put some constraints on the atmospheric 
models of the planet when combining our detection with previous secondary eclipse 
observations of WASP-19b (\citealt{anderson10}, \citealt{gibson10}). Figure 6 shows 
a modified diagram from \citet{burdaj11} which overplots the previously obtained 
$H$- and $K$-band measurements onto various atmospheric models of WASP-19b.
We have also included our estimated $z$'-band eclipse depth in this diagram. Our 
secondary eclipse observation lies somewhat above the simple black-body 
approximation, the closest model being the $A_B$=0.0 (Bond albedo - 
a measure of how much energy is reflected and how much is turned into heat), 
non-black body model, in agreement with the previous $H$- and $K$-band measurements. 
Our data probes a distinctly different regime as far as the planetary atmospheric characteristics
are concerned (note the logarithmic scale on the y-axis). 

\section{Summary}
We present a detection of the secondary eclipse of the extrasolar 
planet WASP-19b in the Sloan $z$' band using the ULTRACAM CCD camera
on the NTT telescope. 
After fitting the transit parameters, we determine a decrease in flux due to the
planet passing behind the star to be 0.088$\pm$0.018\%.
Analysis of the dominant systematic effects in our data give us confidence that the
feature which occurs at the predicted eclipse times is a real feature of the target. 
However, given the limitations of ground-based
observations, especially at optical wavelengths, further observations at similar wavelengths 
will be of great benefit in further constraining the precise eclipse depth and $T_0$ for WASP-19b. 
In addition to this, the near-infrared remains a region over which we have yet to fully explore in 
regard to atmospheric characterisation, and observations such as these will allow for the atmosphere 
at depth to be probed for hot-Jupiters.

We have demonstrated in this paper that meaningful results for ground-based secondary eclipse detections 
can be achieved provided that sufficient care and attention is given to ruling out 
systematic variations and correlations between the flux from the target, and other 
observable/physical parameters.

\acknowledgments

\noindent{Acknowledgements --}

\noindent JB is funded by the Northern Ireland Department of Employment and learning,
and in addition would like to acknowledge the support of the RMBC Queen Elizabeth 
Foundation. SPL acknowledges the support of an RCUK fellowship. CAW, VSD, TRM, SPL 
and ULTRACAM are supported by STFC. We would also like to thank the referee for providing
comments and suggestions which helped improve the clarity and conciseness of the paper. 

{\it Facilities:} \facility{NTT (ULTRACAM)}



\clearpage
\section{Appendix}
\vspace{-10pt}
\begin{figure}[!h]
\centering
\subfloat[Part 1][]{\includegraphics[trim=0cm 0cm 0cm 0cm, clip=true,width=3.2in]{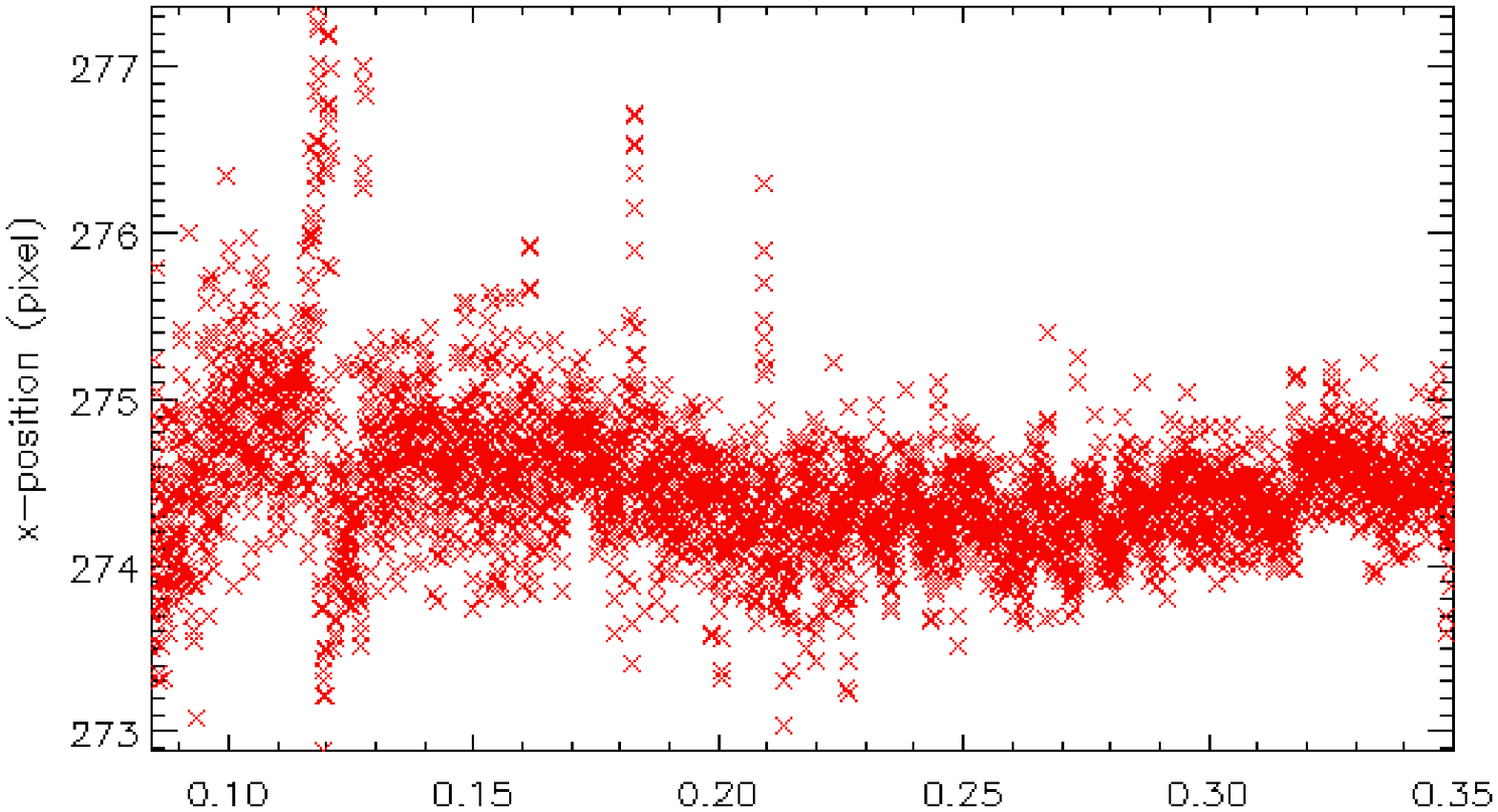} \label{fig:sixfigures-a}}
\subfloat[Part 2][]{\includegraphics[trim=0cm 0cm 0cm 0cm, clip=true,width=3.15in]{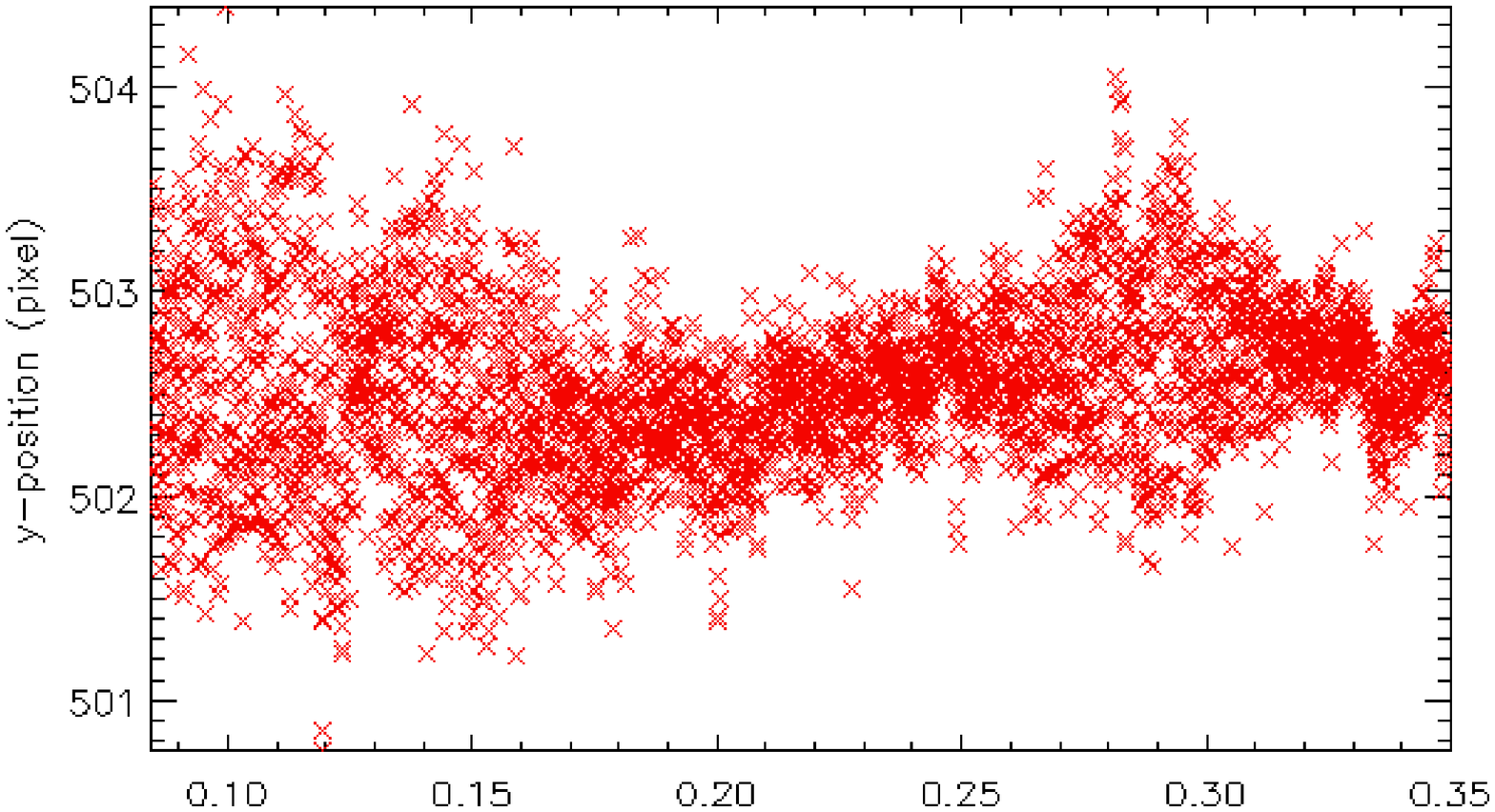} \label{fig:sixfigures-b}}\\
\subfloat[Part 3][]{\includegraphics[trim=0cm 0cm 0cm 0cm, clip=true,width=3.2in]{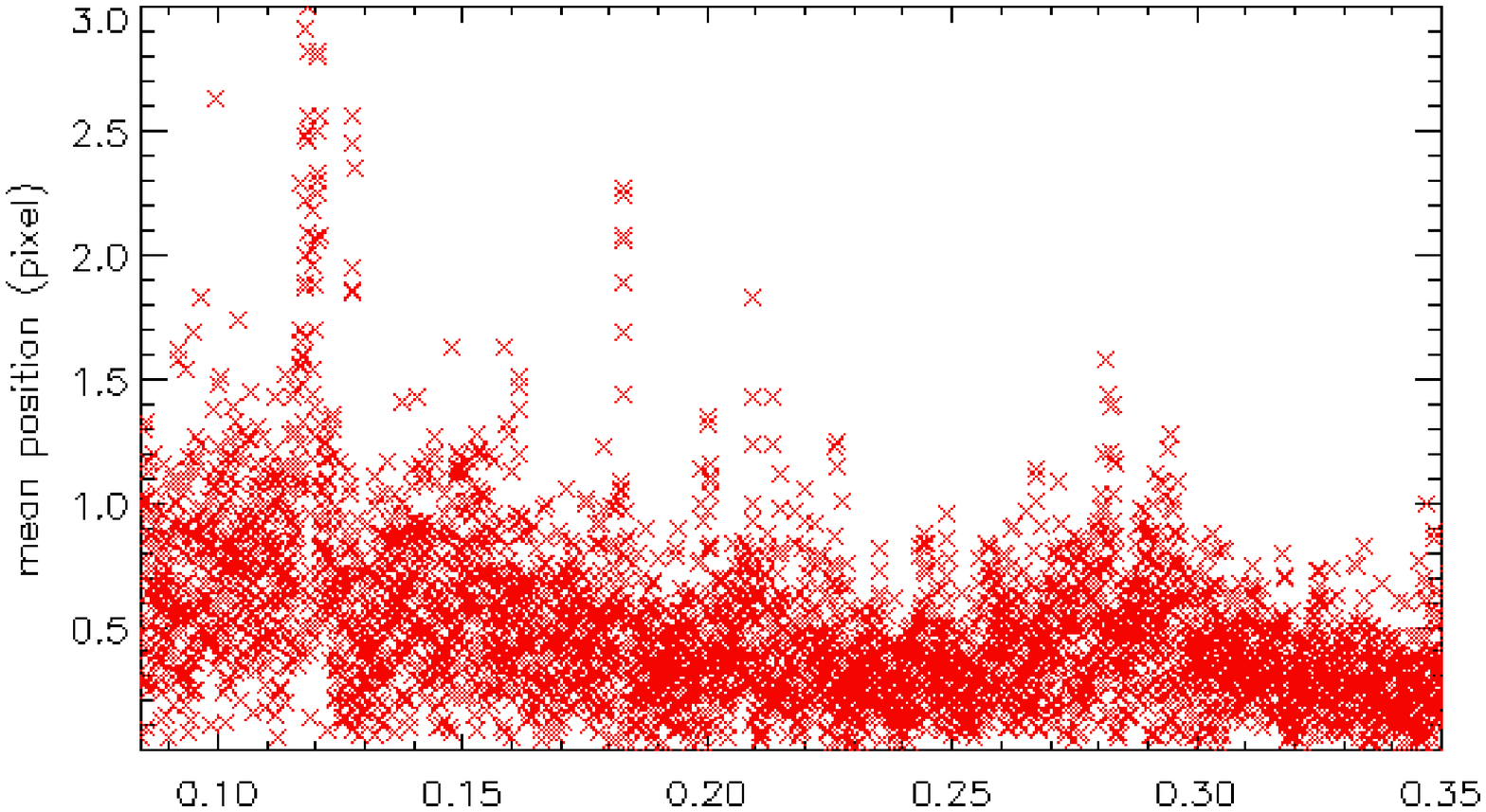} \label{fig:sixfigures-c}}
\subfloat[Part 4][]{\includegraphics[trim=0.5cm 0cm 0cm 0cm, clip=true,width=3.12in]{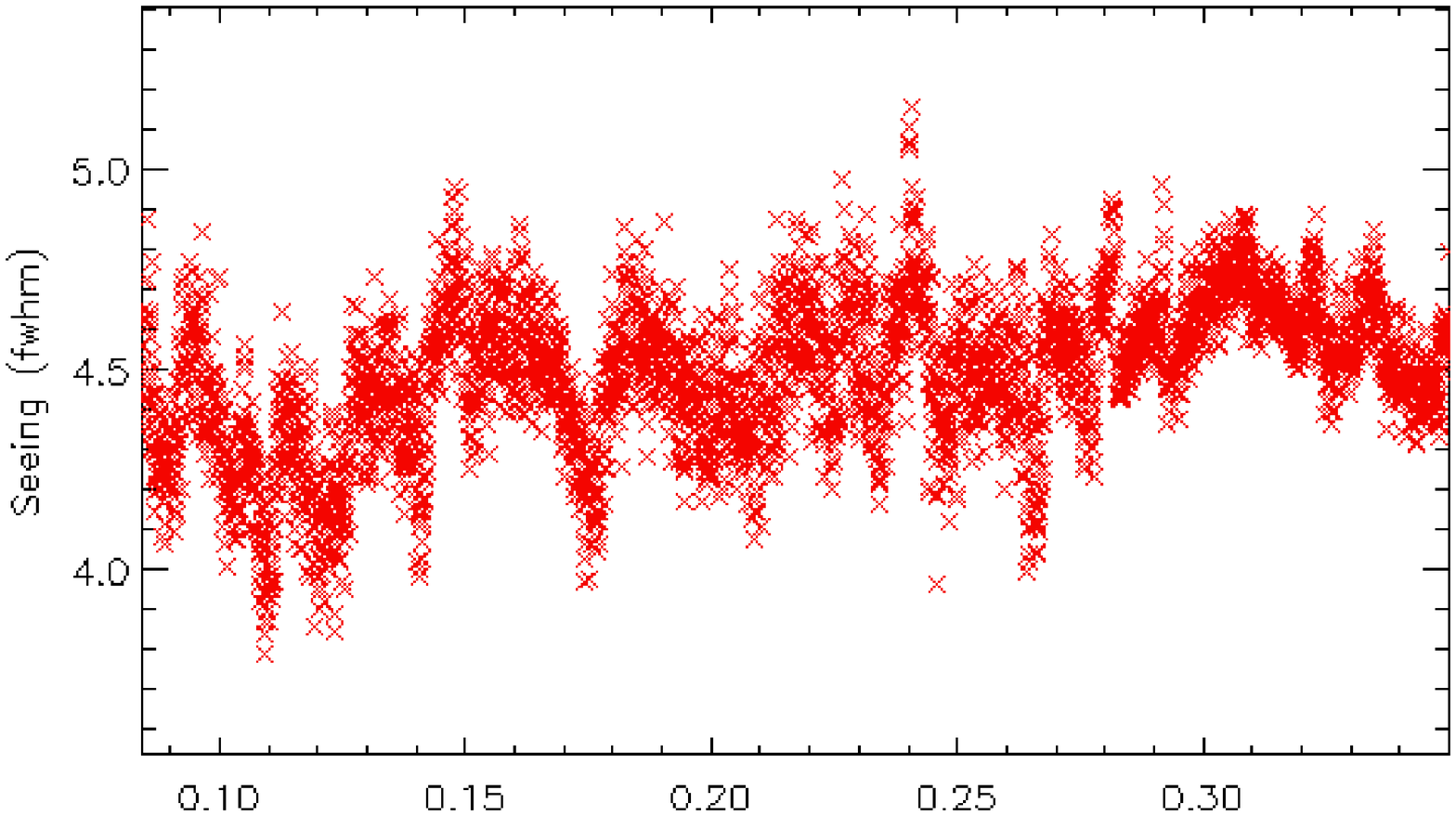} \label{fig:sixfigures-d}}\\
\subfloat[Part 5][]{\includegraphics[trim=0cm 0cm 0cm 0cm, clip=true,width=3.2in]{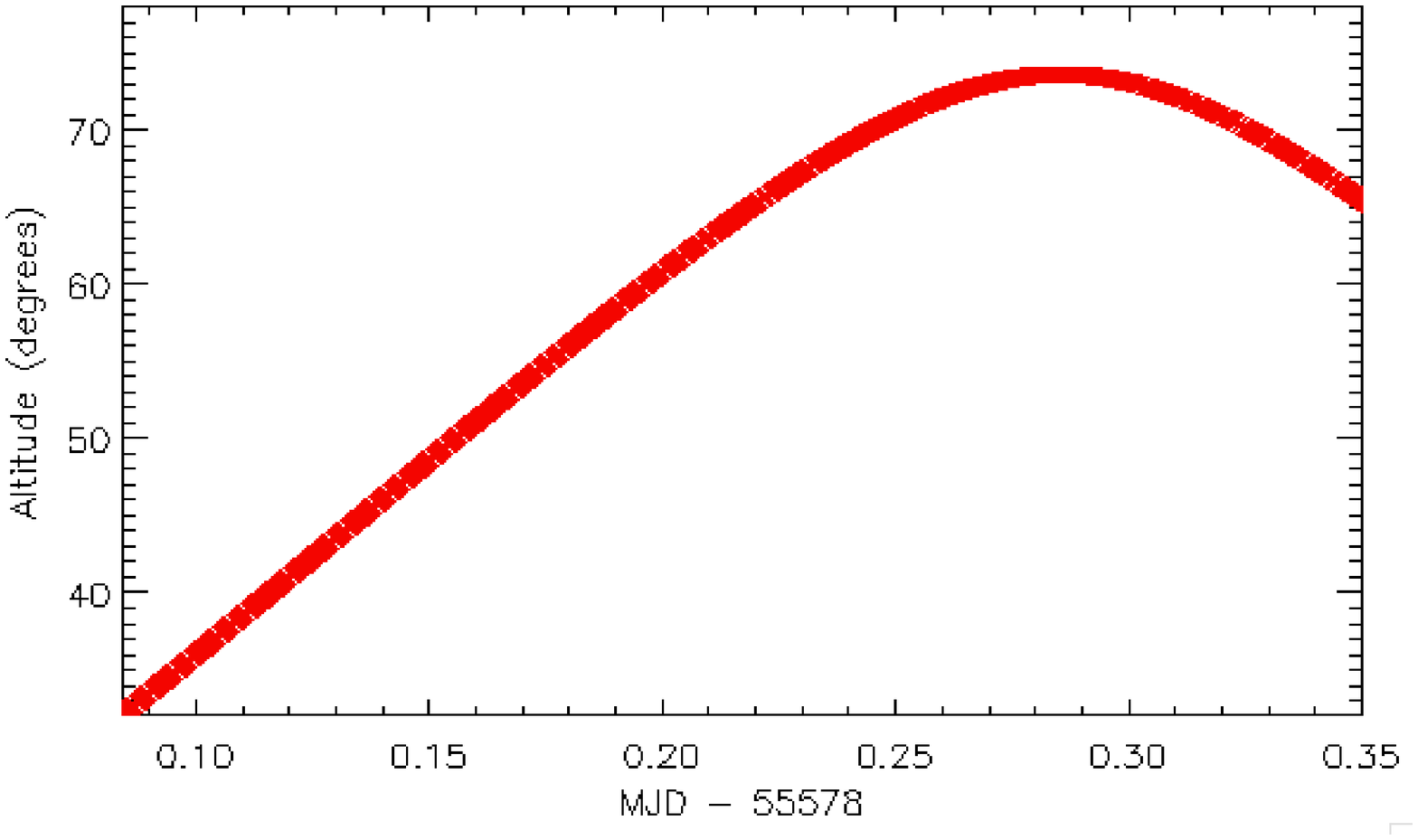} \label{fig:sixfigures-e}}
\subfloat[Part 6][]{\includegraphics[trim=0.5cm 0cm 0cm 0cm, clip=true,width=3.2in]{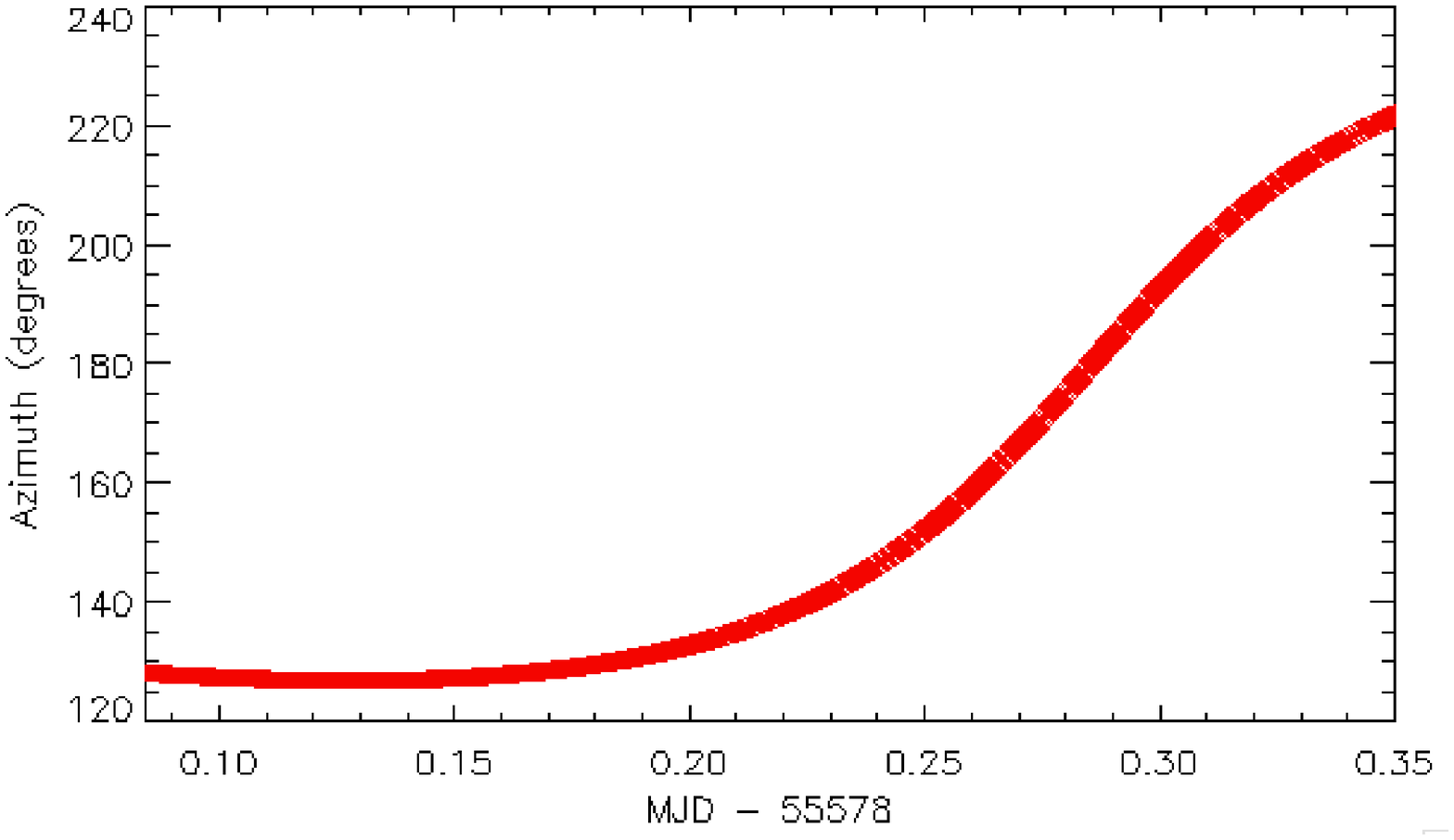} \label{fig:sixfigures-f}}\
\caption{$z$'-band time-series of (a) x position, (b) y position, (c) mean position, (d) seeing, (e) altitude (airmass), (f) azimuth over the course 
of the observations.}
\label{fig:sixfigures}
\end{figure}

\begin{figure}[!h]
\centering
\subfloat[Part 1][]{\includegraphics[trim=0cm 0cm 0.75cm 0cm, clip=true,width=3.3225in]{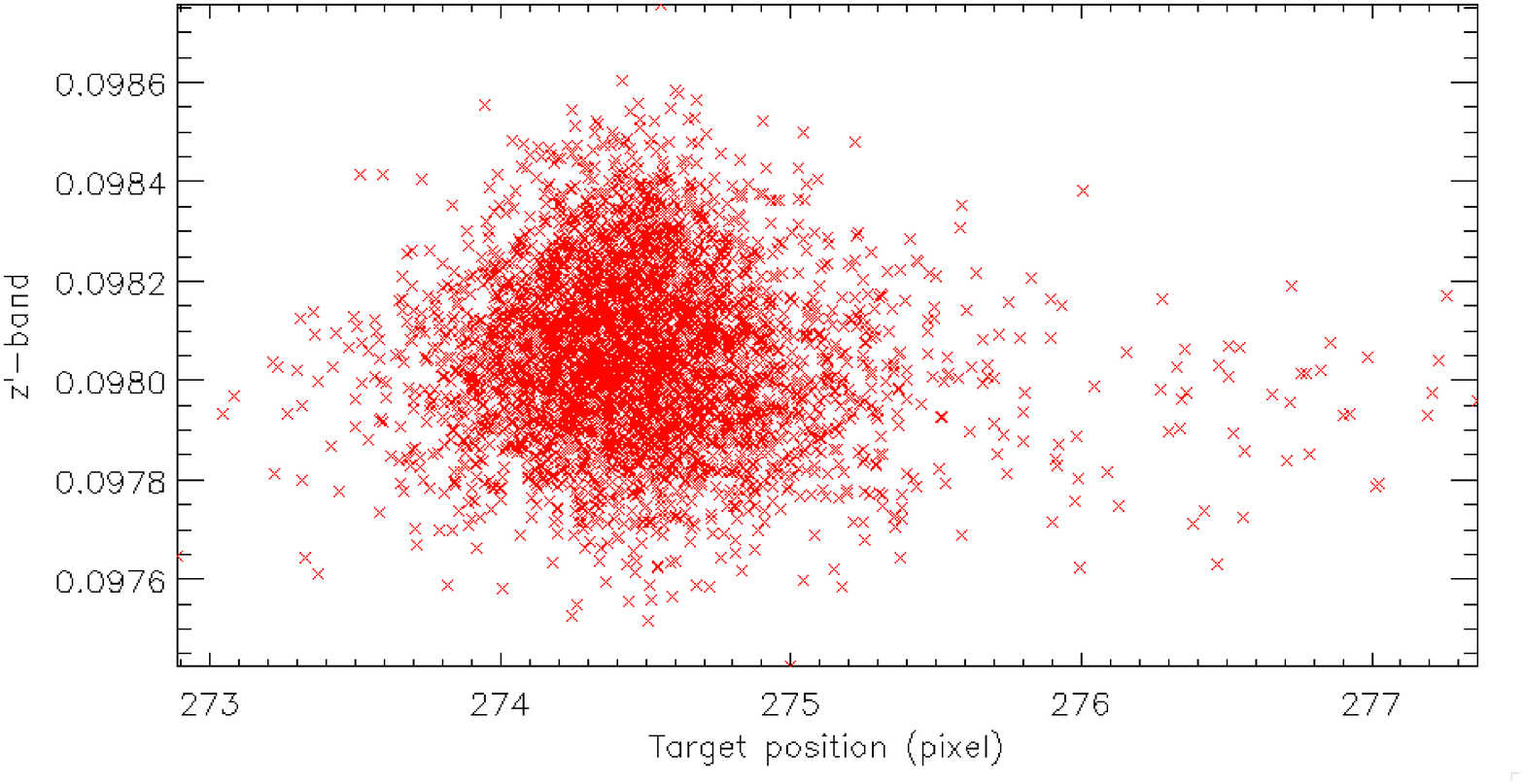} \label{fig:sixfigures-a}}
\subfloat[Part 2][]{\includegraphics[trim=4cm 0cm 0cm 0cm, clip=true,width=3in]{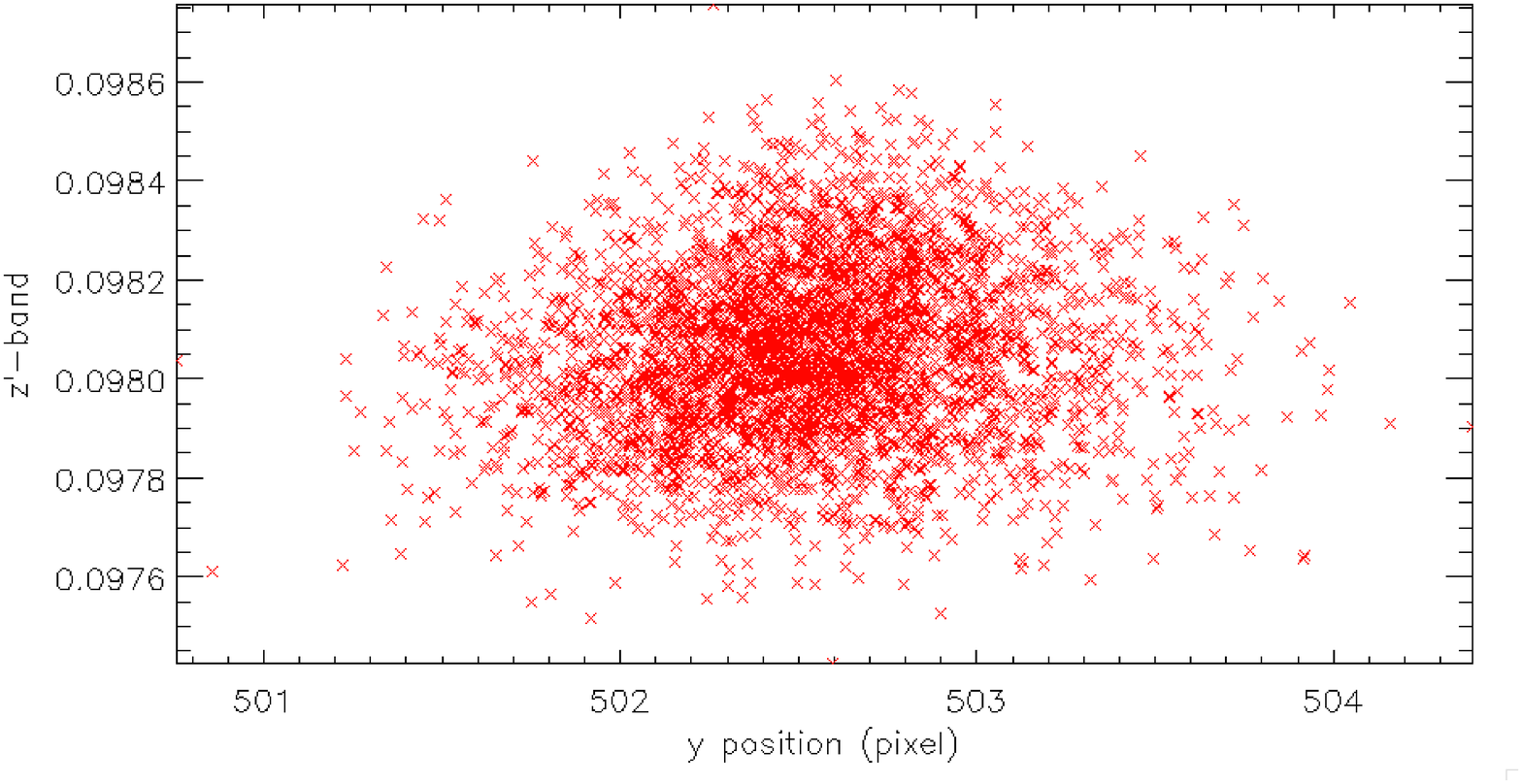} \label{fig:sixfigures-b}}\\
\subfloat[Part 3][]{\includegraphics[trim=0cm 0cm 0.75cm 0cm, clip=true,width=3.37in]{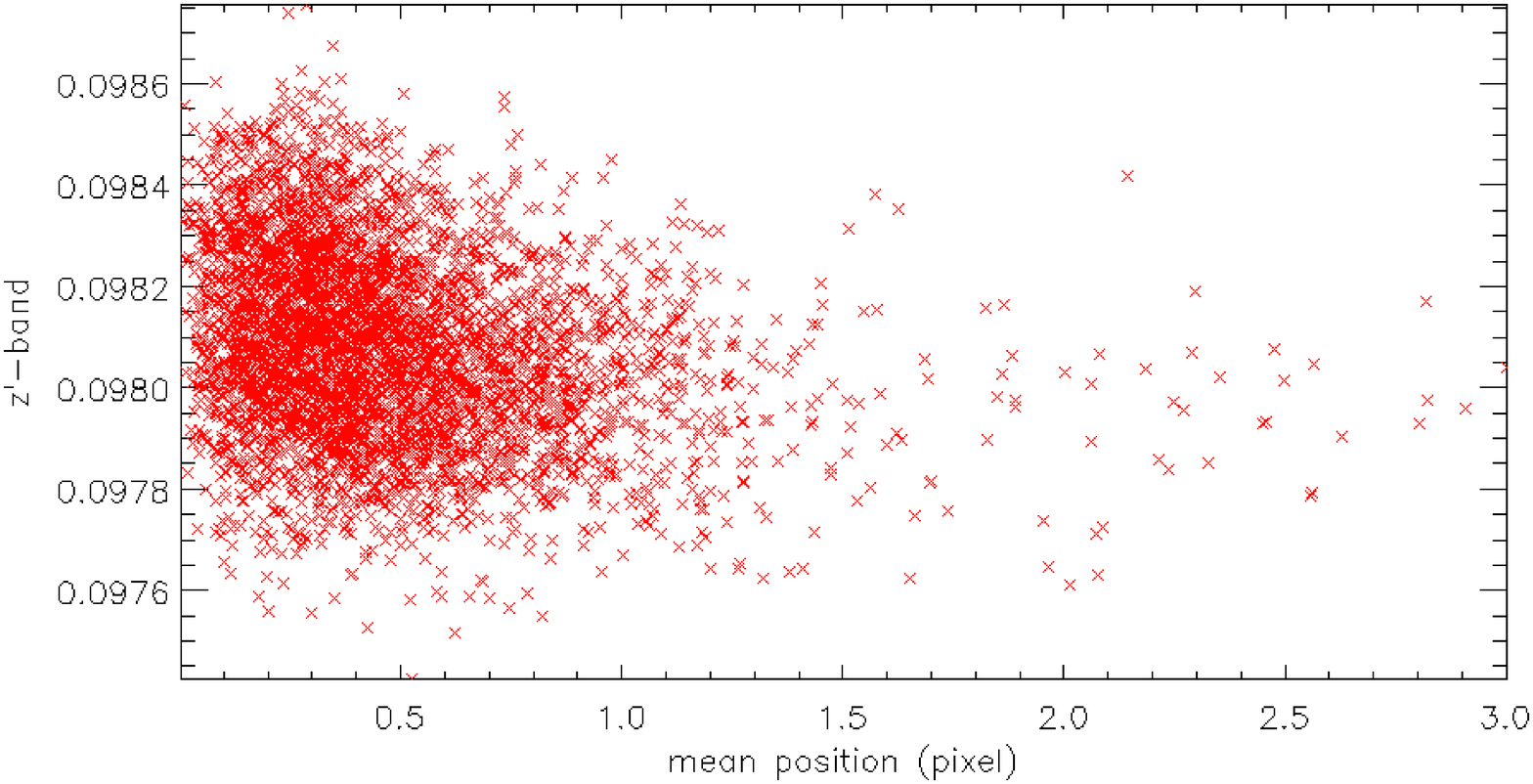} \label{fig:sixfigures-c}}
\subfloat[Part 4][]{\includegraphics[trim=4cm 0cm 0cm 0cm, clip=true,width=3in]{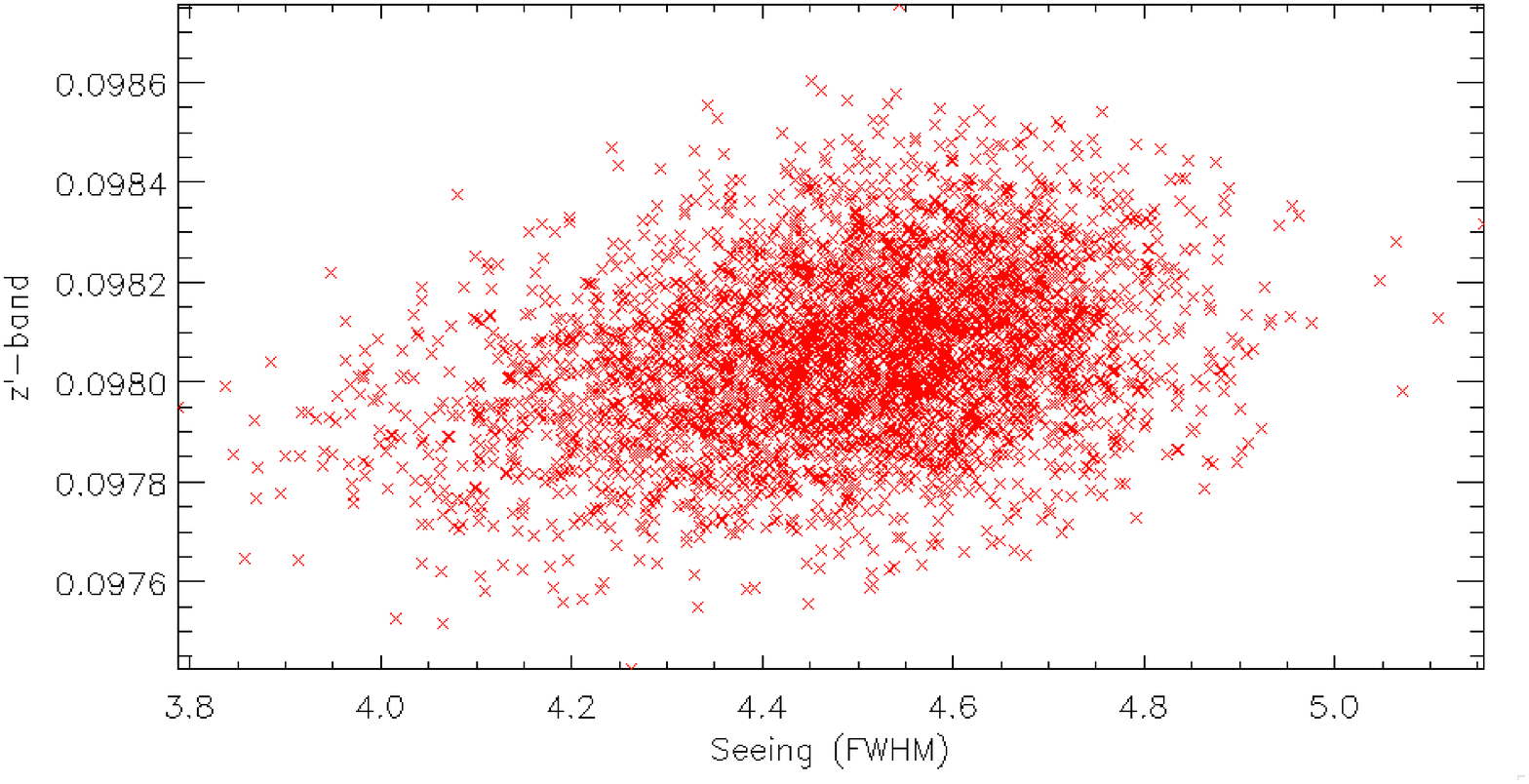} \label{fig:sixfigures-d}}\\
\subfloat[Part 5][]{\includegraphics[trim=0cm 0cm 0.75cm 0cm, clip=true,width=3.35in]{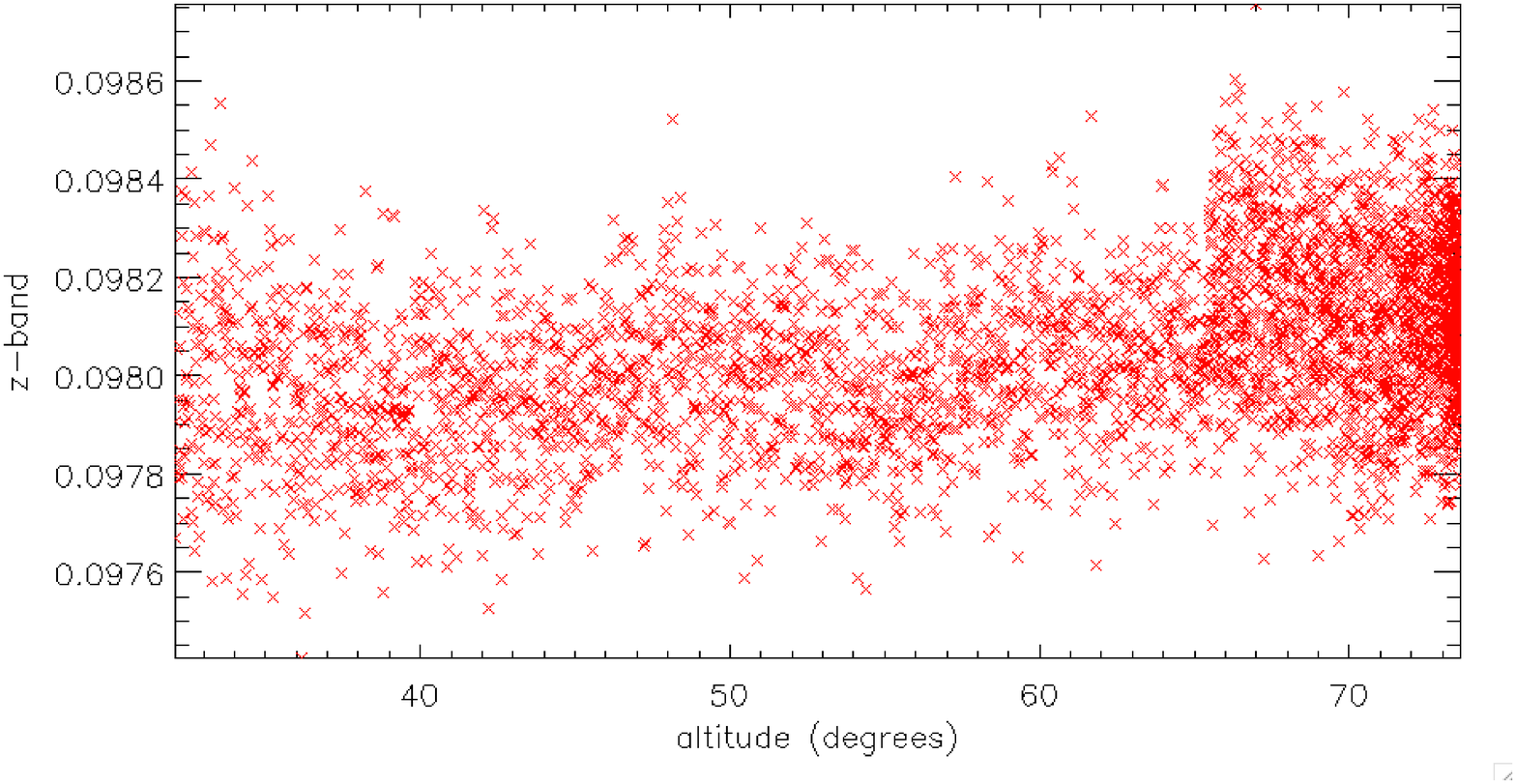} \label{fig:sixfigures-e}}
\subfloat[Part 6][]{\includegraphics[trim=4cm 0cm 0cm 0cm, clip=true,width=3in]{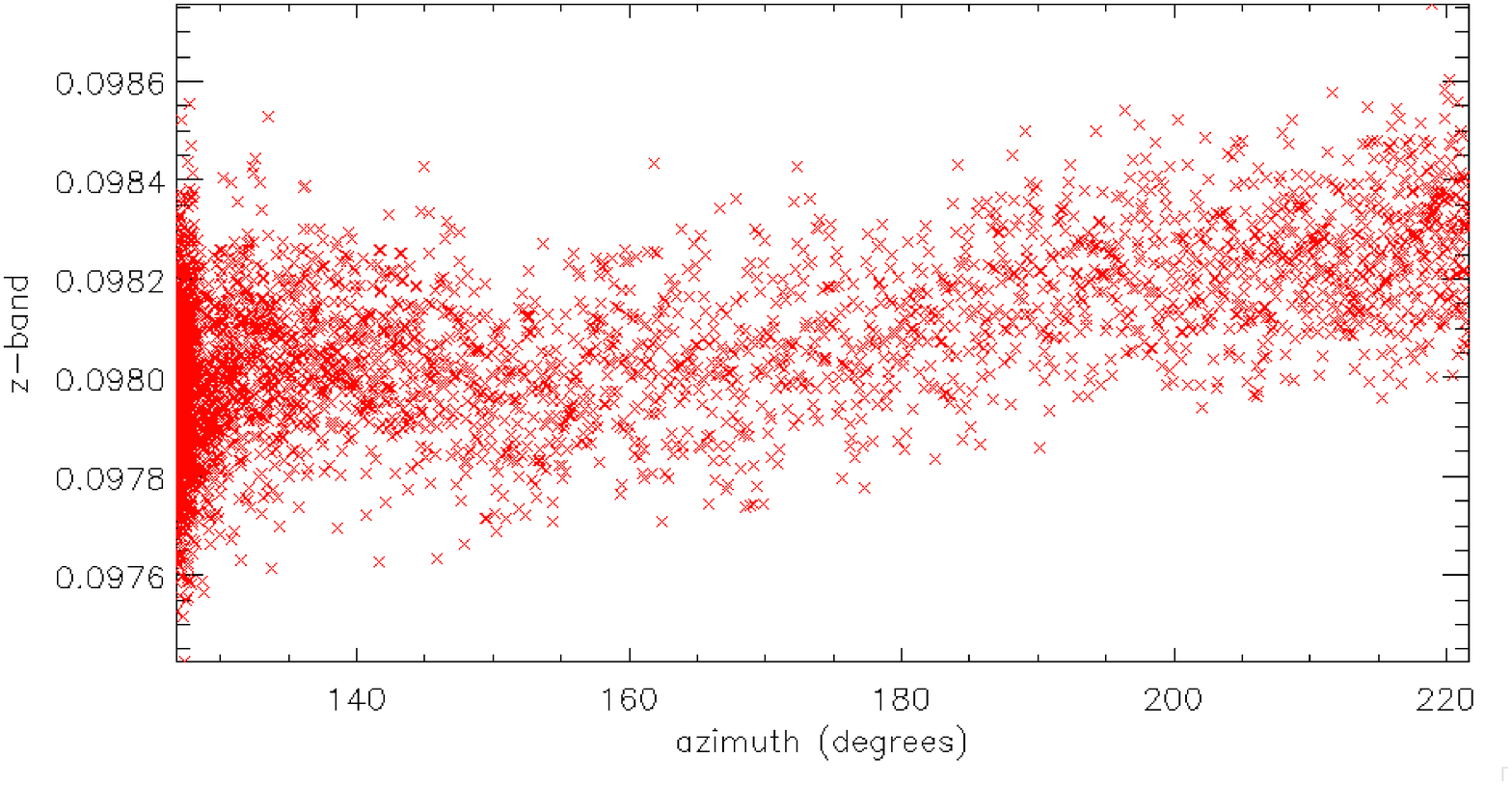} \label{fig:sixfigures-f}}\
\caption{$z$'-band versus (a) x position. (b) y position. (c) mean position. (d) seeing. (e) altitude (airmass). (f) azimuth. Note
that the target position on the chip only exhibits typically sub-pixel motion.}
\label{fig:sixfigures}
\end{figure}

\begin{figure}[!h]
\epsscale{0.9}
\plotone{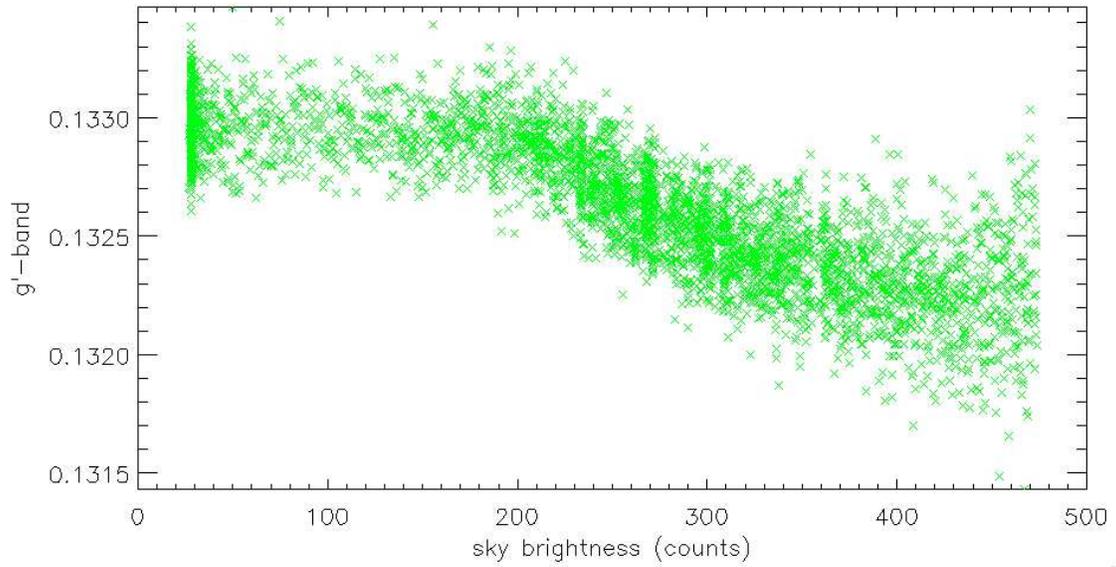}
\caption{$g$'-band vs sky. The correlation here follows the same 2-component trend as the 
$z'$-band vs. sky background, but since the sky background is brighter relative to the target 
in the $g$'-band, this is unsurprising. Again, as with the $z$'-band data, a number of aperture sizes were trialled in order to
assure no flux from the target was leaking into the sky aperture. \label{fig9}}
\end{figure}

\begin{figure}[!h]
\centering
\subfloat[Part 1][]{\includegraphics[trim=0cm 0cm 0.75cm 0cm, clip=true,width=3.3325in]{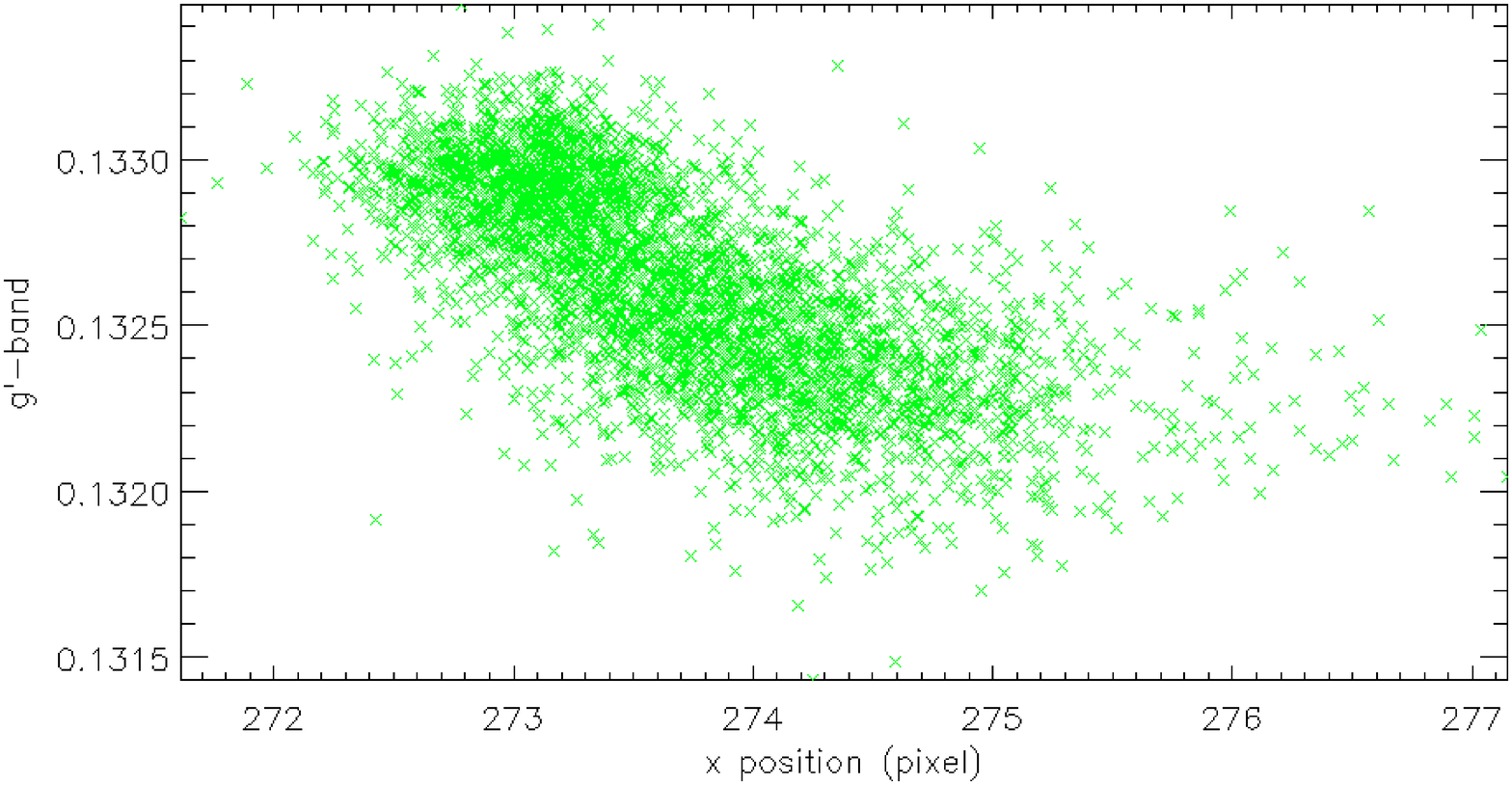} \label{fig:sixfigures2-a}}
\subfloat[Part 2][]{\includegraphics[trim=4cm 0cm 0cm 0cm, clip=true,width=3in]{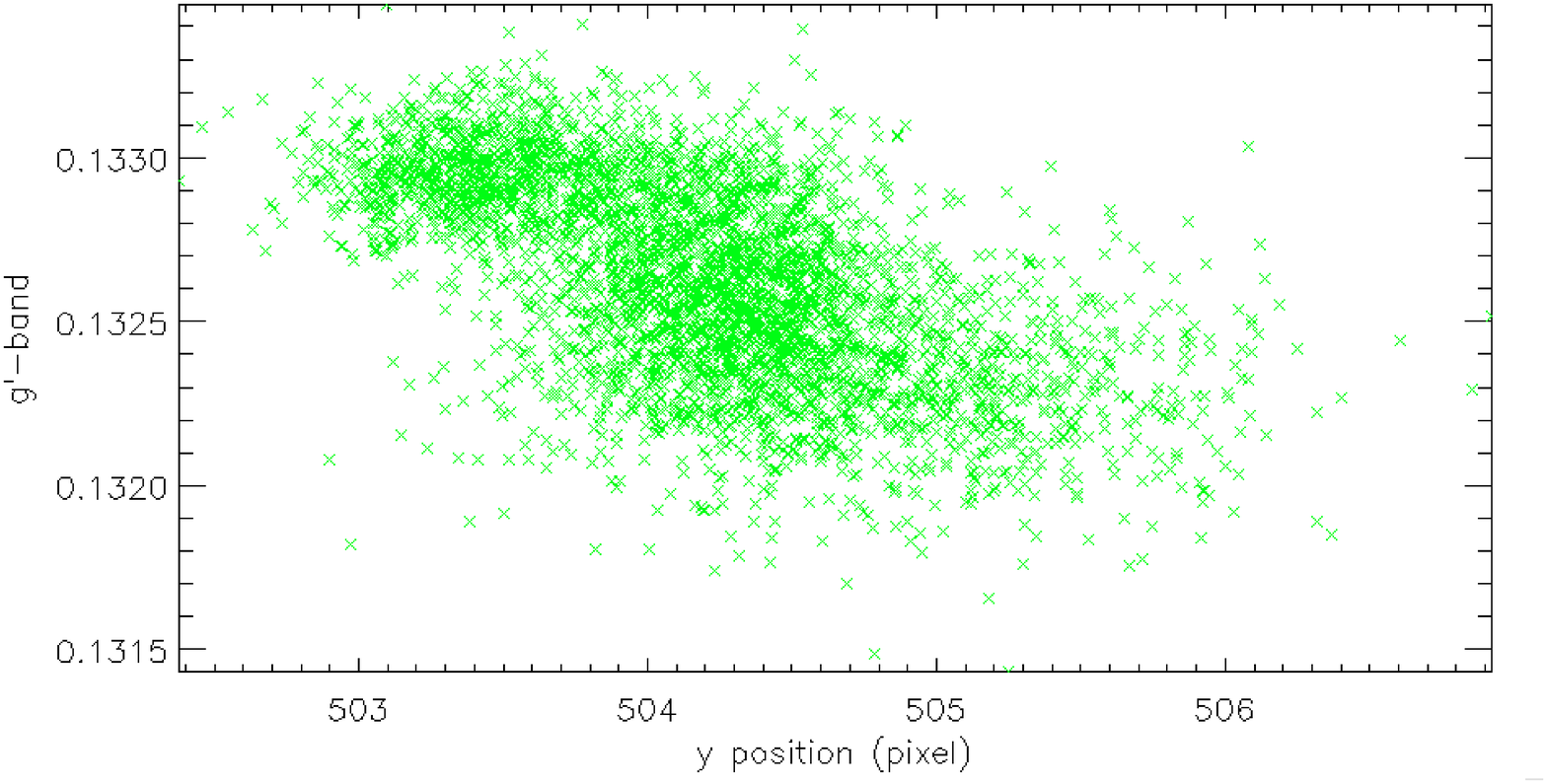} \label{fig:sixfigures2-b}}\\
\subfloat[Part 3][]{\includegraphics[trim=0cm 0cm 0.75cm 0cm, clip=true,width=3.335in]{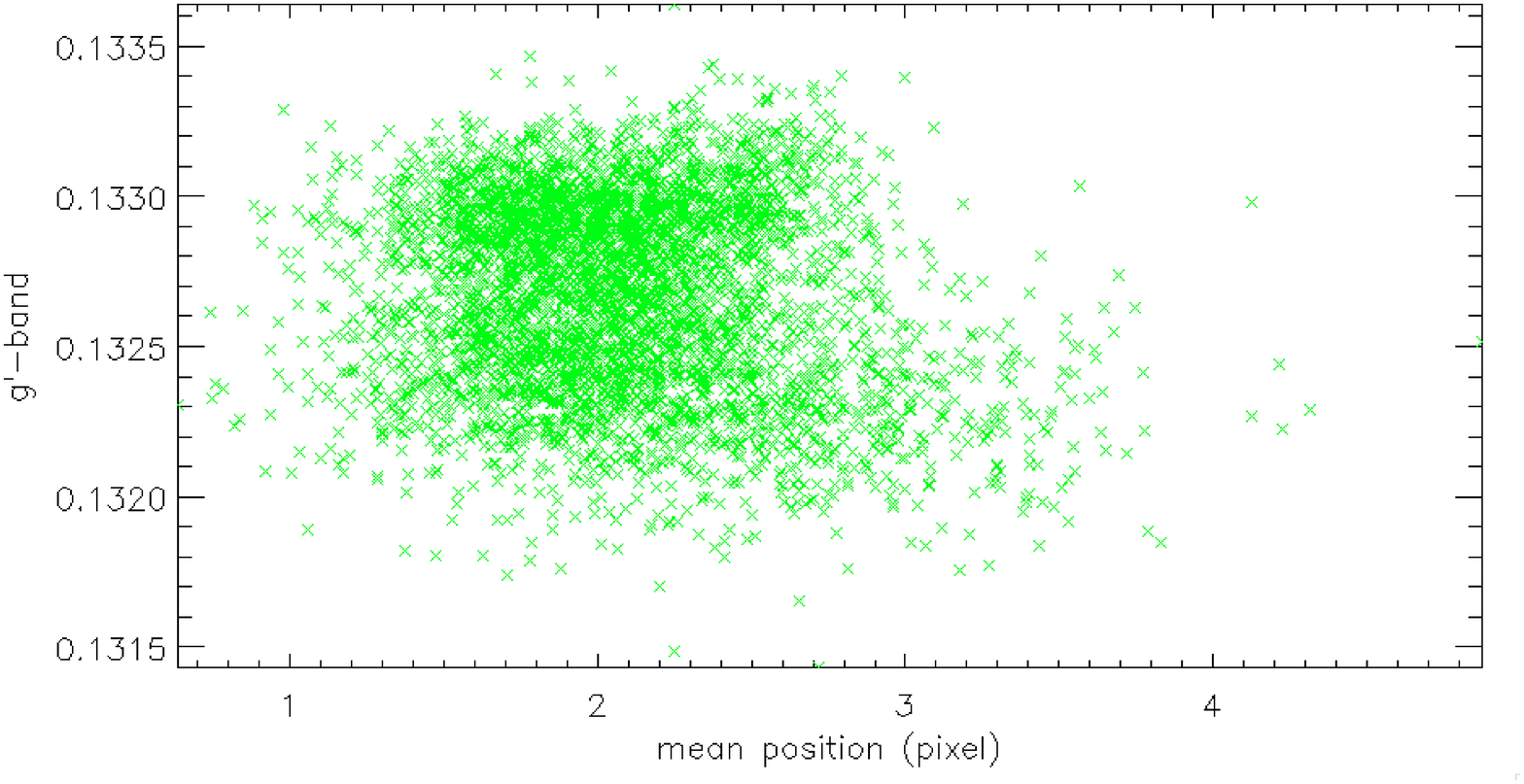} \label{fig:sixfigures2-c}}
\subfloat[Part 4][]{\includegraphics[trim=4cm 0cm 0cm 0cm, clip=true,width=3in]{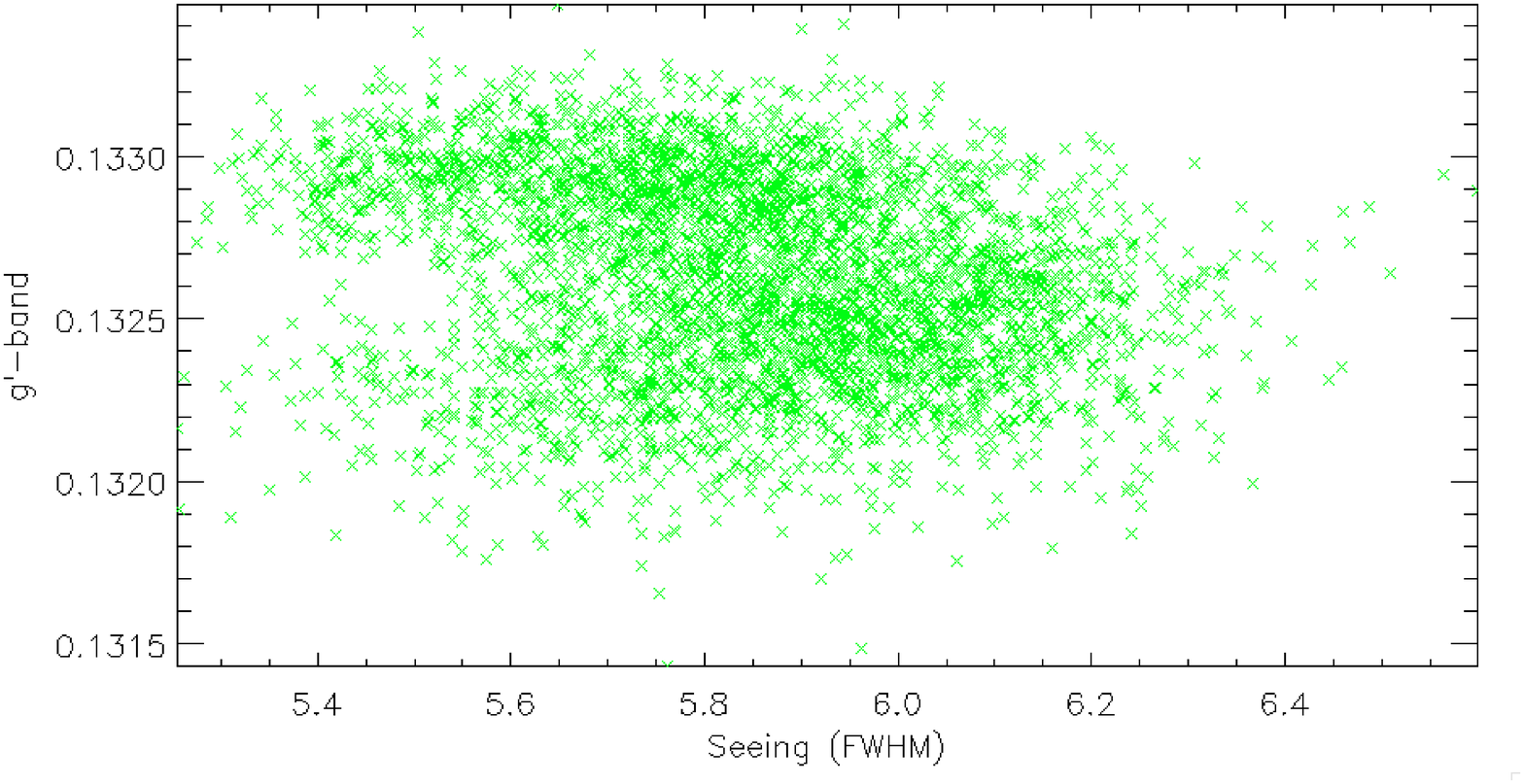} \label{fig:sixfigures2-d}}\\
\subfloat[Part 5][]{\includegraphics[trim=0cm 0cm 0.5cm 0cm, clip=true,width=3.32in]{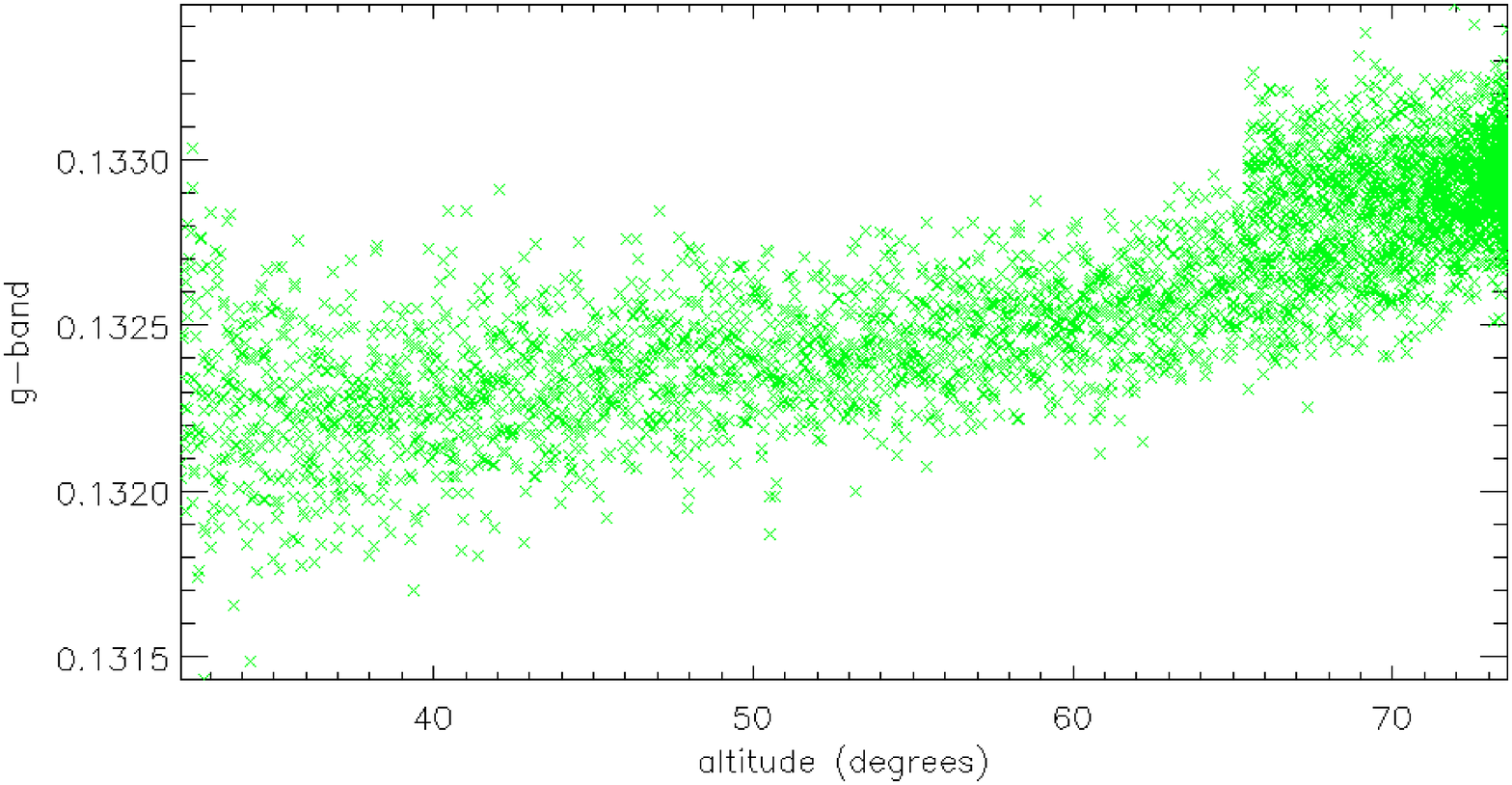} \label{fig:sixfigures2-e}}
\subfloat[Part 6][]{\includegraphics[trim=4cm 0cm 0cm 0cm, clip=true,width=3in]{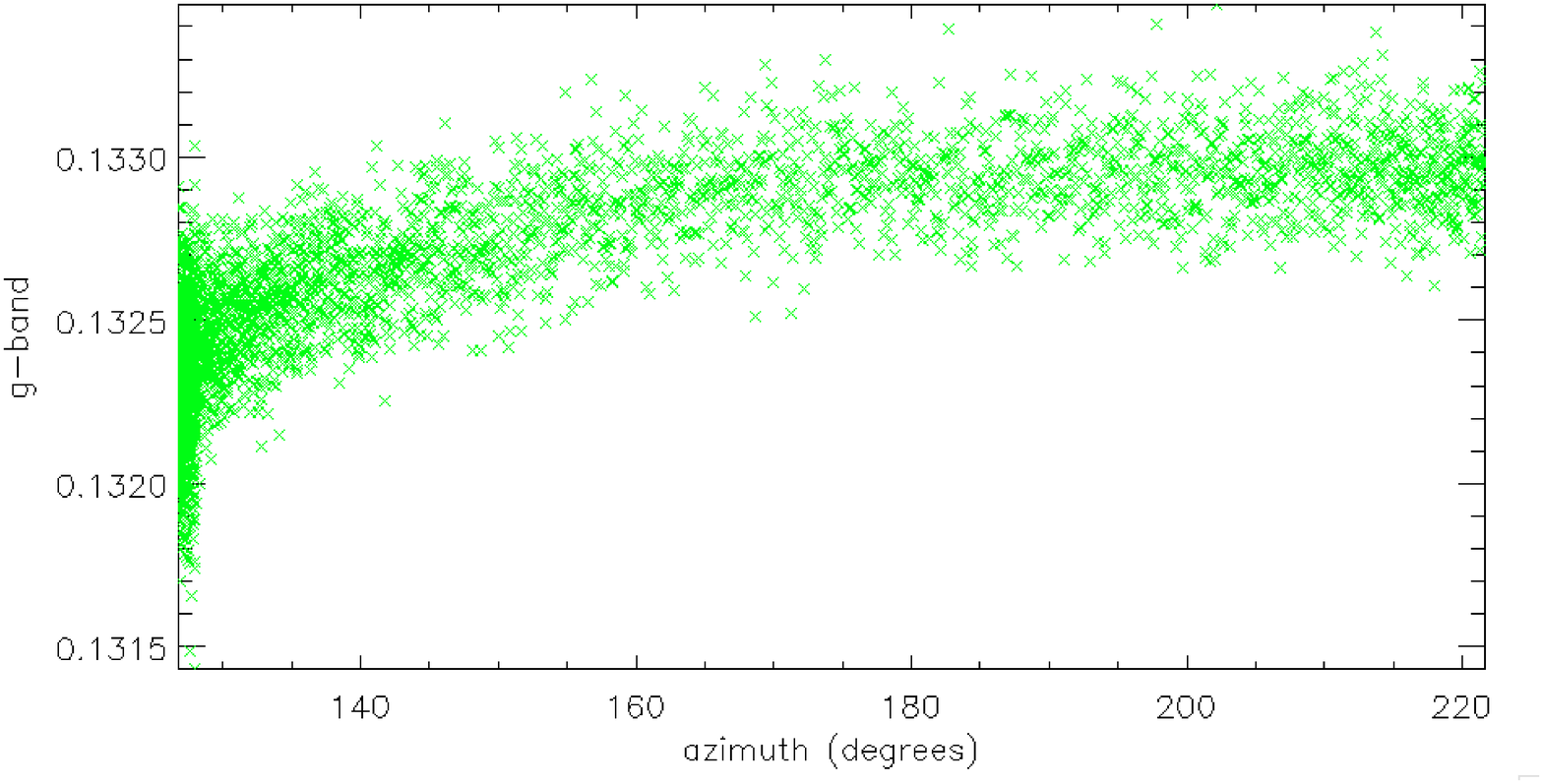} \label{fig:sixfigures2-f}}\
\caption{$g$'-band versus (a) x position. (b) y position. (c) mean position. (d) seeing. (e) altitude (airmass). (f) azimuth. Note that once the sky background (figure 9) is removed, these correlations are significantly reduced, indicating sky background is the dominant source of correlations for the data set.}
\label{fig:sixfigures2}
\end{figure}

\clearpage

\end{document}